# Precision annealing Monte Carlo methods for statistical data assimilation and machine learning

Zheng Fang,[1,*,†] Adrian S. Wong,[1,†] Kangbo Hao,[1,†] Alexander J. A. Ty,[1] and Henry D. I. Abarbanel[1,2]
[1]*Department of Physics, University of California, San Diego, La Jolla, California 92093, USA*
[2]*Marine Physical Laboratory (Scripps Institution of Oceanography), University of California, San Diego, La Jolla, California 92093, USA*

In statistical data assimilation (SDA) and supervised machine learning (ML), we wish to transfer information from observations to a model of the processes underlying those observations. For SDA, the model consists of a set of differential equations that describe the dynamics of a physical system. For ML, the model is usually constructed using other strategies. In this paper, we develop a systematic formulation based on Monte Carlo sampling to achieve such information transfer. Following the derivation of an appropriate target distribution, we present the formulation based on the standard Metropolis-Hasting (MH) procedure and the Hamiltonian Monte Carlo (HMC) method for performing the high-dimensional integrals that appear. To the extensive literature on MH and HMC, we add (1) an annealing method using a hyperparameter that governs the precision of the model to identify and explore the highest probability regions of phase space dominating those integrals, and (2) a strategy for initializing the state-space search. The efficacy of the proposed formulation is demonstrated using a nonlinear dynamical model with chaotic solutions widely used in geophysics.

## I. INTRODUCTION

Two seemingly distinct challenges for systematically transferring information from a well-curated (but noisy) data set to a model of the processes producing the data, namely statistical data assimilation (SDA) [1–6] and machine learning [7–14], have been shown to be equivalent in their formal structure [7]. In artificial neural networks, the rules that direct the activities from layer to layer are equivalent to the rules for the temporal development of dynamical models in statistical data assimilation. In SDA, the number of measurements at each observation time plays the same role as the number of independent input-output pairs in machine learning.

In this paper we formulate the information transfer problem as a Monte Carlo evaluation of a high-dimensional expected value integral. This formulation can be applied both to physical dynamical systems and to machine-learning problems [7,15]. We also explore the evaluation of such integrals and add to the Monte Carlo procedures a strategy to identify the dominant contribution to the expected values. We use a problem description close to SDA in physical sciences [1], nevertheless, the statements and lessons identified here are directly usable in machine learning [7].

We first establish the problem in Sec. II and then briefly discuss the Hamiltonian Monte Carlo (HMC) method [16–18] in Sec. III. HMC overcomes the difficulties in traditional Monte Carlo by complementing the original state space with a set of canonical variables moving in "time." Here, in particular, we study HMC with two additional features that are specified in Sec. V.

(1) We use a *precision annealing* (PA) method derived from the formulation of the "action" $A(\mathbf{X}) = -\log[\pi(\mathbf{X}\,|\,\mathbf{Y})]$. The target distribution $\pi(\mathbf{X}\,|\,\mathbf{Y})$, conditioned on all observations, governs the expected values of the model state variables and parameters.

(2) We use a different way to initialize the Monte Carlo search.

We call the combination of these strategies *precision annealing Hamilton Monte Carlo* (PAHMC).

We also report some results from standard Metropolis-Hastings (MH) Monte Carlo calculations [19,20] in which the proposals for movements in the state space are based on random perturbations of the present location. We label these methods as random proposal (RP) searches. We employ our PA and initialization strategies in both the RP and the HMC analyses.

To show the effectiveness of our formulation, both approaches are demonstrated on a chaotic dynamical model widely used in geophysics [21]. We present in Sec. VI results on estimating the observed and unobserved state variables of the model, estimating the unknown model parameters, as well as predicting forward in time.

### A. Example physics problems addressed by the methods discussed

Here we very briefly discuss two physics problems where we have applied *earlier* methods for data assimilation:

(i) determining the electrophysiological properties of isolated neurons from the HVC nucleus of an avian birdsong system, and

*Corresponding author: zfang@physics.ucsd.edu
†These authors contributed equally to this work.

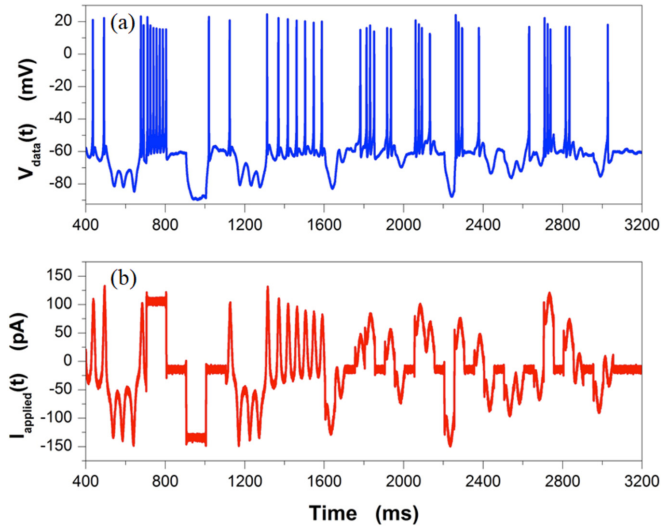

FIG. 1. An illustrative example of the data assimilation challenge. This figure shows data from D. Margoliash of the University of Chicago and D. Meliza of the University of Virginia. An interneuron within the nucleus HVC of the avian brain was isolated in a glass dish in the laboratory in an *in vitro* experiment. An electrode was inserted into the body of the neuron and the applied current $I_{app}(t)$ shown in red in (b) was injected into the neuron. The resulting membrane potential response, shown in blue in (a), was measured using the same electrode. From these data, one is asked to estimate all the parameters in the model, the three (unobserved) grating variables $A_j(t)$, and $V(t)$ (as it is noisy) in Eq. (1).

(ii) estimating the number of required observations in the shallow water equations, a core component of numerical weather models.

These particular biophysics or geophysics problems are discussed in detail in earlier works, and we do not dwell on the details of SDA in their solution. We do sketch the data assimilation issues in each, and we consider them as natural arenas for the use of the SDA methods discussed in this paper. We have not yet applied PAHMC to these problems.

### *1. Hodgkin-Huxley models for individual neurons in the HVC nucleus of the avian song system*

The biophysical equations describing the dynamics of neurons were established by work done before and after World War II. Hodgkin and Huxley [22–24] were two of the researchers whose names are prominent in understanding that equations imposing current conservation on the ions flowing into and departing from the neuron body (soma) and equations capturing the voltage-dependent permeability of the cell membrane to these ions would provide a quantitative framework for the biophysical description of the neural processes (see Fig. 1).

In the experiments they performed, they considered two ion channels for sodium and potassium ions flowing through proteins penetrating the cell membrane. They also introduced "leak" channel describing other aspects of neuron behavior.

The equations they proposed, and tested, have the form

$$
\begin{aligned}
C \frac{dV(t)}{dt} &= g_{Na} m(t)^3 h(t) [E_{Na} - V(t)] \\
&\quad + g_K n(t)^4 [E_K - V(t)] + g_L [E_L - V(t)] \\
&\quad + I_{DC} + I_{app}(t).
\end{aligned} \quad (1)
$$

The three voltage-dependent "gating" variables, i.e., $A_j(t) = m(t), h(t), n(t)$, $0 \leqslant A_j(t) \leqslant 1$, are taken to satisfy the first-order kinetics

$$\frac{dA_j(t)}{dt} = \frac{A_{j0}(V(t)) - A_j(t)}{\tau_j(V(t))}. \quad (2)$$

The parameters $g_{Na}$, $g_K$, and $g_L$ in the voltage equation (1) are constants, while the gating variables are state variables. $A_{j0}(V(t))$ and $\tau_j(V(t))$ are voltage-dependent functions. The first is dimensionless and sets the scale for the gating variables, and the second is a voltage-dependent timescale for the activity of the gating variables.

This is a $(D = 4)$-dimensional dynamical system that has rich behavior [22–24]. In laboratory experiments one can directly measure the cross membrane voltage $V(t)$, but no instruments are available to measure the gating variables. In the general language used below, this means that the three state variables $A_j(t)$ are *unobserved*.

An experiment consists of selecting a stimulating current $I_{app}(t)$, which is typically known. With only $V(t)$ observed, the challenge is to estimate all the fixed parameters in Eq. (1) as well as the $A_j(t)$ over an observation interval $[t_0, t_{final}]$. We also estimate $V(t)$ over this interval as it is always noisy.

Once this is done, the model for the neuron is complete, and with a set of initial conditions at $t = t_{final}$, i.e., $\{V(t_{final}), A_j(t_{final})\}$, we may solve this initial value problem, with fixed parameters and known $I_{app}(t)$ for $t \geqslant t_{final}$. Validation (or not) of the model associated with the observations comes from solving Eqs. (1) and (2) for $t \geqslant t_{final}$, and comparing the model output $V(t)$ with observed values. In the validation interval, no information about the observations for $V(t)$, $t \geqslant t_{final}$, are used to assimilate additional information into the model.

A published example of the methods used before this paper can be found in Ref. [25].

Why should we expect this will work? Because the model is nonlinear in the state variables, they are generically coupled together through the model. Information is passed from the stored data through the observable $V(t)$ and determines (the unobserved) $A_j(t)$, consistent with the data.

### *2. Shallow water equations on a β plane*

As the depth of the atmosphere/ocean fluid layer (order 10–15 km) is markedly less than the earth's radius, the shallow water equations for two-dimensional flow are an excellent approximation to the fluid dynamics in such a geometry [26,27]. Three fields on a mid-latitude plane describe the fluid flow: the east-west velocity $u(\mathbf{r}, t)$, the north-south velocity $v(\mathbf{r}, t)$, and the height of the fluid $h(\mathbf{r}, t)$, with $\mathbf{r} = \{x, y\}$. The fluid is taken as a single, constant density layer and is driven by wind stress $\tau(\mathbf{r}, t)$ at the surface $z = 0$ through an Ekman layer. These physical processes satisfy the following dynamical equations with $\mathbf{u}(\mathbf{r}, t) \equiv \{u(\mathbf{r}, t), v(\mathbf{r}, t)\}$:

$$\frac{\partial \mathbf{u}(\mathbf{r}, t)}{\partial t} = -\mathbf{u}(\mathbf{r}, t) \cdot \nabla \mathbf{u}(\mathbf{r}, t) - g \nabla h(\mathbf{r}, t)$$
$$+ \mathbf{u}(\mathbf{r}, t) \times f(y) \hat{\mathbf{z}} + A \nabla^2 \mathbf{u}(\mathbf{r}, t) - \epsilon \mathbf{u}(\mathbf{r}, t),$$
$$\frac{\partial h(\mathbf{r}, t)}{\partial t} = -\nabla \cdot [h(\mathbf{r}, t) \mathbf{u}(\mathbf{r}, t)] - \hat{\mathbf{z}} \cdot \mathrm{curl}\left[\frac{\tau(\mathbf{r}, t)}{f(y)}\right]. \quad (3)$$

The Coriolis force is linearized about the equator $f(y) = f_0 + \beta y$ and the wind-stress profile is selected to be $\tau(\mathbf{r}) = \{F \cos(2\pi y)/\rho, 0\}$. The parameter $A$ represents the viscosity in the shallow water layer, $\epsilon$ is Rayleigh friction, and $\hat{\mathbf{z}}$ is the unit vector in the $z$ direction. With the chosen fixed parameters, the shallow water flow is chaotic, and the largest Lyapunov exponent for this flow is $\lambda_{\max} = 0.0325$ h$^{-1}$ $\approx$ $1/31$ h$^{-1}$.

We analyzed this flow using the enstrophy-conserving discretization scheme from Ref. [28] on a periodic grid of size $N^2$ for $N = 16, 32, 64$ with periodic boundary conditions. Given a grid on which to solve the fluid dynamical equations, we generated $3N^2$ values of $\{u(x, y, t), v(x, y, t), h(x, y, t)\}$ on the $(x, y)$ grid. $\Delta t = 36$ s was taken to be the time step for the solution.

We performed a "twin experiment" where we solved Eq. (3) on a grid and added noise to each of the $D = 3N^2$ output time series. These are stored away as our data. A subset of these data, an $L$-dimensional time series, were presented to the model using a "nudging" data assimilation method [29]. We estimated that approximately 70% of the $D = 3N^2$ degrees of freedom must be observed in order to synchronize the model output with the data. These results on the required number of observations agree with previous calculations using random proposal Monte Carlo Methods, presented in Ref. [30].

As the results are consistent across the various grid sizes that were investigated, we restricted our discussion here to the case in which $N = 16$ and $L = 0.68 D \approx 524$, with $L$ being the dimension of the observed time series. We are confident that despite the numerical challenges associated with scaling the algorithm up to larger $D$, the results presented here for $N = 16$ also remain valid for higher grid resolution.

The number $L = 524$ comes from asking when the data assimilation method used (not PAHMC) permitted accurate prediction beyond the observation window (in time). In the twin experiments, we knew all $D = 768$ time series on the grid, so we could compare the solutions to Eq. (3) to the estimations from the data assimilation procedure. Often, for $L \leqslant 524$ the estimations differed significantly in the observation window as well as in the prediction window. When $L \geqslant 524$, this did not happen.

## II. SDA AND MACHINE LEARNING

We begin with a description of an SDA framework within which we will discuss the transfer of information to a dynamical model of the processes producing the observations. This sets the notation and frames our subsequent discussions.

Within an observation window $[t_0, t_{\mathrm{final}}]$, we make a set of measurements at times $t = \{\tau_0, \ldots, \tau_F\}$ where $t_0 \leqslant$ $\tau_0 < \tau_F \leqslant t_{\mathrm{final}}$. At each of these measurement times $k = 0, \ldots, F$, we observe an $L$-dimensional vector $\mathbf{y}(\tau_k) = (y_1(\tau_k), \ldots, y_L(\tau_k))$. The number of observations $L$ at each measurement time $t = \tau_k$ is typically less, sometimes much less, than the number of degrees of freedom $D$ in the model of the observed system, i.e., $L \leqslant D$. These data are stored in a library and otherwise unaltered during the remainder of the discussion. We designate the collection of observations made from time $t = \tau_0$ up to $\tau_F$ as $\mathbf{Y} \equiv \{\mathbf{y}(\tau_0), \ldots, \mathbf{y}(\tau_F)\}$.

Our quantitative characterization of the dynamical processes that produced these data is through the chosen model. In SDA, our choices are based on knowledge of the physical dynamics assumed to be describing the data source.

If we are using the structures here to perform a task in machine learning, the model may be divorced from knowledge of physics and constructed through statistical principles. The model describes the interactions among and the evolution of the states of the target system. From the data $\mathbf{Y}$, we want to estimate the *observed* and *unobserved* states of the model as a function of time as well as to estimate any time-independent physical parameters in the model. At the end of the observation window $t = t_{\mathrm{final}}$, we use all the estimated values of model states and parameters to predict the model response to new forcing of the system for $t \geqslant t_{\mathrm{final}}$. The predictions are then used to validate (or not) the model selected by us.

We call the $D$-dimensional state of the model $\mathbf{x}(t) = (x_1(t), \ldots, x_D(t))$. The model is constructed to describe the dynamical behavior of the observations through a set of differential equations in continuous time and is written as

$$\frac{dx_a(t)}{dt} = \mathcal{F}_a(\mathbf{x}(t), \boldsymbol{\theta}), \quad a = 1, \ldots, D. \quad (4)$$

In discrete time with equal intervals $\Delta t$, $t_m = t_0 + m\Delta t$, where $m = 0, \ldots, M$ and $t_M = t_{\mathrm{final}}$, the dynamics can be written as $x_a(t_{m+1}) = f_a(\mathbf{x}(t_m), \boldsymbol{\theta})$ or as

$$x_a(m + 1) = f_a(\mathbf{x}(m), \boldsymbol{\theta}), \quad (5)$$

where $\boldsymbol{\theta}$ is a set of time-independent parameters $\boldsymbol{\theta} = \{\theta_1, \ldots, \theta_{N_p}\}$ associated with the model. The discrete-time model $f_a(\mathbf{x}(m), \boldsymbol{\theta})$ is related to its continuous counterpart $\mathcal{F}_a(\mathbf{x}(t), \boldsymbol{\theta})$ via the choice for solving the continuous time flow for $\mathbf{x}(t)$ through a discrete-time numerical solution method [31].

We work henceforth in discrete time. For convenience we choose the observation times $\tau_k$ to be integer multiples of $\Delta t$ as well,

$$\tau_k = t_0 + n_k \Delta t, \quad k = 0, \ldots, F \quad (6)$$

where $\{n_0, \ldots, n_F\}$ is a set of integers and $\{n_k\} \leqslant M$.

Starting from the initiation of the observation window at $t_0$, we must use our model equations to move the state variables $\mathbf{x}(0)$ from $t_0$ to $\tau_1 = t_0 + n_1 \Delta t$, where the first measurement is made. Then, we use the model dynamics again to move along to $\tau_2 = t_0 + n_2 \Delta t$, where the second measurement is made, and so forth until we reach the time of the last measurement $\tau_F = t_0 + n_F \Delta t$ and finally move the model from $\mathbf{x}(\tau_F)$ to $\mathbf{x}(t_{\mathrm{final}}) = \mathbf{x}(M)$.

We collect all the $D$-dimensional vectors $\mathbf{x}(m)$ for $m = 0, \ldots, M$ along with all of the $N_p$ parameters into what we call the *path* $\mathbf{X}$, $\mathbf{X} \equiv \{\mathbf{x}(0), \ldots, \mathbf{x}(M); \theta_1, \ldots, \theta_{N_p}\}$. The dimension of the path is $(M+1)D + N_p$. A visual representation of this discussion is given in Fig. 2.

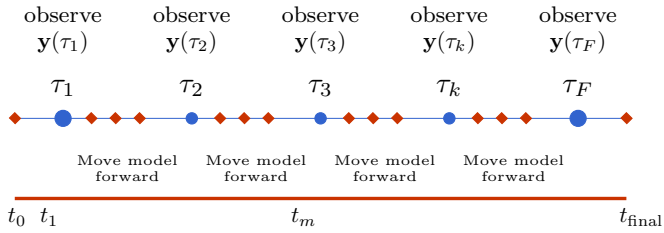

FIG. 2. A visual representation of the time window $t_0 \leqslant t \leqslant t_{\text{final}}$ during which $L$-dimensional ($L \leqslant D$) observations $\mathbf{y}(\tau_k)$ are made at times $\tau_k = t_0 + n_k \Delta t$ for $k = 0, \ldots, F$. We also show the times at which the $D$-dimensional nonlinear model $\mathbf{x}(m+1) = \mathbf{f}(\mathbf{x}(m), \boldsymbol{\theta})$ is used to move forward from time $t_m$ to $t_{m+1}$ where $t_m = t_0 + m\Delta t$, $m = 0, 1, \ldots, M$, and $t_{\text{final}} = t_M$.

We need three ingredients to affect our transfer of information from the collection of all measurements $\mathbf{Y}$ to the model $\mathbf{x}(m+1) = \mathbf{f}(\mathbf{x}(m), \boldsymbol{\theta})$ along the path $\mathbf{X}$ through the observation window $[t_0, t_{\text{final}}]$:

(1) our noisy data $\mathbf{Y}$,

(2) a model of the processes producing $\mathbf{Y}$ (this model is devised by our experience and knowledge of those processes), and

(3) the third ingredient is comprised of methods to achieve the transfer of information from $\mathbf{Y}$ to properties of the model (we would like to estimate all the components in $\mathbf{X}$; this will be our focus from here on).

In this paper we describe and implement identical test problems on two well-explored methods of Monte Carlo search:

(1) *The original method* [19,20]. This starts at some location $\mathbf{X}_i$ in path space, makes proposals to move to a candidate location $\mathbf{X}_c$ by sampling a modification to $\mathbf{X}_i$ drawn from a symmetric distribution. It then accepts or rejects $\mathbf{X}_c$ using the Metropolis-Hastings rule. This procedure is also outlined in Sec. III A. Under some conditions [31,32], this will explore the target distribution $\pi(\mathbf{X} | \mathbf{Y})$ well and allow the approximate evaluation of expected values of functions $\langle G(\mathbf{X}) \rangle = \int d\mathbf{X}\, G(\mathbf{X}) \pi(\mathbf{X} | \mathbf{Y})$. We label this very well-explored construction as the "random proposal" (RP) Monte Carlo method.

(2) *The Hamiltonian Monte Carlo method* [16–18]. HMC takes off in another direction. It proposes moves from $\mathbf{X}_i$ in an enlarged space reached by adding canonical momenta $\mathbf{P}$ to $\mathbf{X}$. The proposals are made using Hamilton's equations of classical mechanics. HMC is reviewed in Sec. III. A numerical example is given in Sec. IV. The computational complexity of HMC is discussed in Sec. VII.

If the transfer of information is successful and, according to some metric of success, we arrange matters so that at the measurement times $\tau_k$, the $L$ model variables $\mathbf{x}(t = \tau_k)$ associated with $\mathbf{y}(\tau_k)$ are approximately equal to each other, we are *not* finished yet.

At this point, we have only demonstrated that the model is consistent with the known data $\mathbf{Y}$. We must further use the model, completed by the estimates of $\boldsymbol{\theta}$ and the full state of the model at the final time $\mathbf{x}(M)$, to predict forward for $t > t_M$, and we should succeed in comparison with measurements for $\mathbf{y}(\tau)$ for $\tau > t_M$. As the measure of success, we may use the same metric as used in the observation window. No further information from the observations is passed to the model in the prediction window.

The same overall setup also applies to artificial neural networks in supervised machine learning [8,9], where the training set is used to transfer information into the model (by adjusting the weights); the test set is used to make predictions and validate the model. The process of prediction is called generalization.

### A. The data are noisy, the model has errors

It is inevitable that the data we collect are noisy and the model we construct to describe the dynamics has errors. This means we must address the conditional probability distribution $\pi(\mathbf{X} | \mathbf{Y})$ along the process of transferring information from $\mathbf{Y}$ to the model.

Assume for the moment that we have an equal number of observation and model time steps, so $n_k = k$ in Eq. (6) for $k = 1, \ldots, M$. We now define the shorthand notation: $\mathbf{Y}_0^k \equiv \{\mathbf{y}(0), \ldots, \mathbf{y}(k)\}$ and $\mathbf{X}_0^k \equiv \{\mathbf{x}(0), \ldots, \mathbf{x}(k); \boldsymbol{\theta}\}$ for $k \leqslant M$.

From $t = \tau_k$ to $\tau_{k+1}$ we have a recursion relation relating $\pi(\mathbf{X}_0^{k+1} | \mathbf{Y}_0^{k+1})$ to $\pi(\mathbf{X}_0^k | \mathbf{Y}_0^k)$:

$$\pi(\mathbf{X}_0^{k+1} | \mathbf{Y}_0^{k+1}) = \frac{\pi(\mathbf{x}(k+1), \mathbf{y}(k+1), \mathbf{X}_0^k, \mathbf{Y}_0^k)}{\pi(\mathbf{y}(k+1), \mathbf{Y}_0^k) \cdot \pi(\mathbf{x}(k+1), \mathbf{X}_0^k, \mathbf{Y}_0^k)}$$
$$\cdot \pi(\mathbf{x}(k+1) | \mathbf{x}(k)) \cdot \pi(\mathbf{X}_0^k, \mathbf{Y}_0^k)$$
$$= \frac{\pi(\mathbf{y}(k+1), \mathbf{X}_0^{k+1} | \mathbf{Y}_0^k)}{\pi(\mathbf{y}(k+1) | \mathbf{Y}_0^k) \cdot \pi(\mathbf{X}_0^{k+1} | \mathbf{Y}_0^k)}$$
$$\cdot \pi(\mathbf{x}(k+1) | \mathbf{x}(k)) \cdot \pi(\mathbf{X}_0^k | \mathbf{Y}_0^k)$$
$$= \exp\left[\text{CMI}(\mathbf{y}(k+1), \{\mathbf{x}(k+1), \mathbf{X}_0^k\} | \mathbf{Y}_0^k)\right]$$
$$\cdot \pi(\mathbf{x}(k+1) | \mathbf{x}(k)) \cdot \pi(\mathbf{X}_0^k | \mathbf{Y}_0^k)$$
$$= \frac{\pi(\mathbf{y}(k+1) | \mathbf{x}(k+1), \mathbf{X}_0^k, \mathbf{Y}_0^k)}{\pi(\mathbf{y}(k+1) | \mathbf{Y}_0^k)}$$
$$\cdot \pi(\mathbf{x}(k+1) | \mathbf{x}(k)) \cdot \pi(\mathbf{X}_0^k | \mathbf{Y}_0^k). \quad (7)$$

We have used the Markov property of the model $\pi(\mathbf{x}(k+1) | \mathbf{X}_0^k, \mathbf{Y}_0^k) = \pi(\mathbf{x}(k+1) | \mathbf{x}(k))$ and identified $\text{CMI}(a, b | c) = \log\left[\frac{\pi(a,b|c)}{\pi(a|c)\pi(b|c)}\right]$ which is Shannon's conditional mutual information [33] telling us how many bits (when using $\log_2$) we know about $a$ when observing $b$ conditioned on $c$. For us, $a = \mathbf{y}(k+1)$, $b = \{\mathbf{x}(k+1), \mathbf{X}_0^k\}$, and $c = \mathbf{Y}_0^k$. The appearance of the CMI is an explicit indication that it is information that is transferred from the observations to the model at each measurement time.

Now set aside the condition $n_k = k$, and use Eq. (7) to move backward from the end of the observation window at $t_{\text{final}} = t_0 + M\Delta t$, through the observations at times $\tau_k$, to the start of the window at $t_0$. Up to factors independent of $\mathbf{X}$, we

arrive at

$$\pi(\mathbf{X} \mid \mathbf{Y}) \propto \left[ \prod_{k=0}^{F} \pi\left(\mathbf{y}(\tau_k) \mid \mathbf{x}(\tau_k), \mathbf{Y}_0^{k-1}\right) \right] \\ \times \left[ \prod_{m=1}^{M} \pi(\mathbf{x}(m) \mid \mathbf{x}(m-1)) \right] \pi(\mathbf{x}(0)), \quad (8)$$

where $\mathbf{Y}_0^{-1}$ is the empty set.

$\pi(\mathbf{X} \mid \mathbf{Y}) \propto \exp[-A(\mathbf{X})]$ defines the *action* $A(\mathbf{X})$. Conditional expected values for functions $G(\mathbf{X})$ along the path $\mathbf{X}$ are given by

$$\langle G(\mathbf{X}) \rangle = E[G(\mathbf{X}) \mid \mathbf{Y}] = \frac{\int d\mathbf{X}\, G(\mathbf{X})\, e^{-A(\mathbf{X})}}{\int d\mathbf{X}\, e^{-A(\mathbf{X})}}, \quad (9)$$

where $d\mathbf{X} = \prod_{m=0}^{M} d^D\mathbf{x}(m) \prod_{j=1}^{N_p} d\theta_j$. All factors in the action independent of $\mathbf{X}$ cancel out here.

From Eq. (8), the action reads as

$$A(\mathbf{X}) = -\sum_{k=0}^{F} \log\left[\pi\left(\mathbf{y}(\tau_k) \mid \mathbf{x}(\tau_k), \mathbf{Y}_0^{k-1}\right)\right] \\ -\sum_{m=1}^{M} \log[\pi(\mathbf{x}(m) \mid \mathbf{x}(m-1))] - \log[\pi(\mathbf{x}(0))]. \quad (10)$$

The first sum in Eq. (10) represents the modification to $\pi(\mathbf{X} \mid \mathbf{Y})$ each time an observation $\mathbf{y}(\tau_k)$ is made. The second sum includes the stochastic transitions of the state variables.

The last term includes the distribution of the initial condition of the model. We often have no knowledge about $\pi(\mathbf{x}(0))$, just as we may have no knowledge about the distribution of the parameters $\boldsymbol{\theta}$. In this situation, we take them to be uniform over the dynamical range of the model variables (or the parameters). They then cancel between the numerator and the denominator in Eq. (9).

What $G(\mathbf{X})$ are of interest? A natural choice is the path of the model states and parameters itself: $G(\mathbf{X}) = \mathbf{X}$. Another could be the covariance matrix of $\mathbf{X}$.

The action further simplifies to what we call the "standard model" of SDA when (1) the observations $\mathbf{y}(\tau_k)$ are related to their model counterparts through Gaussian noise with zero mean and a diagonal precision matrix $\mathbf{R}_m = R_m \mathbf{I}$, and (2) the model errors are associated with Gaussian errors of mean zero and a diagonal precision matrix $\mathbf{R}_f = R_f \mathbf{I}$. The standard model action takes the form

$$A(\mathbf{X}) = \sum_{k=0}^{F} \sum_{\ell=1}^{L} \frac{R_m}{2(F+1)} [x_\ell(\tau_k) - y_\ell(\tau_k)]^2 \\ + \sum_{m=0}^{M-1} \sum_{a=1}^{D} \frac{R_f}{2M} [x_a(m+1) - f_a(\mathbf{x}(m), \boldsymbol{\theta})]^2. \quad (11)$$

The first term in Eq. (11) is the *measurement error* and the second term, the *model error*.

One way to explicitly see the balance condition between the measurement error and the model error is to examine the resulting Euler-Lagrange equation when the model time becomes continuous when $\Delta t \to 0$. This is discussed in [7].

### B. Goal of SDA

Our challenge is to perform expected value integrals such as Eq. (9). One should anticipate that the dominant contribution comes from the maxima of $\pi(\mathbf{X} \mid \mathbf{Y})$ or, equivalently, the minima of $A(\mathbf{X})$. Near such minima, the two contributions to the action, the measurement error and the model error, balance each other to accomplish the explicit transfer of information from the data to the model.

When $\mathbf{f}(\mathbf{x}(n), \boldsymbol{\theta})$ is nonlinear in $\mathbf{X}$, the expected value integral (9) is not Gaussian, and we need to approximately evaluate integrals of this form in order to complete the task of transferring information.

Two generally useful approaches for evaluating this kind of high-dimensional integral are Laplace's method [34] and the collection of techniques using Monte Carlo sampling [16,19,20,31,35].

The Laplace-method evaluations of expected value integrals are discussed in [36–39]. They do not sample from $\pi(\mathbf{X} \mid \mathbf{Y})$ away from its maximum. Evaluating corrections to the leading Laplace contributions is familiar as perturbation theory in statistical physics [40]. The convergence of such perturbation methods can depend sensitively on the functional form of the action in $\mathbf{X}$. In addition, the need for repeatedly using the Hessian matrix of $A(\mathbf{X})$ may further hinder their use in practice.

The rest of the paper is organized as follows. In Sec. III, we briefly review the random proposal and the Hamiltonian Monte Carlo methods. We then give a two-dimensional example comparing the two Monte Carlo methods in Sec. IV. Then, in Sec. V, we formulate the proposed method to evaluate Eq. (9). We then present a detailed study on a nonlinear, chaotic dynamical model [21] in Sec. VI as a demonstration. Following the numerical calculations, we discuss in Sec. VII the complexity and scaling properties of PAHMC. We remark on the use of state-space representation and summarize the results of this paper in Sec. VIII.

## III. MONTE CARLO METHODS

### A. Random-proposal Monte Carlo

Performing integrals such as Eq. (9) via Monte Carlo searches requires a method to sample from $\pi(\mathbf{X} \mid \mathbf{Y})$ in order to identify and use those regions in the path space $\mathbf{X}$ yielding the dominant contribution to the integral. The original Monte Carlo procedure devised by Metropolis and Hastings [19,20,41] has been widely explored. It samples from a probability distribution such as $\pi(\mathbf{X} \mid \mathbf{Y})$, seeking the maxima of that distribution. Proposals on how to move about path space are made by perturbing the present location at random. This random-proposal (RP) procedure is in theory ergodic, meaning that the proposals can reach any region in the path space with nonzero probability given a sufficient number of proposed steps [42]. This property makes it an unbiased method for performing the expected value integrals. In practice, the RP procedure often suffers from slow mixing in high-dimensional models [32,43,44].

This method proposes different steps via a (usually symmetric) proposal distribution, from a present path-space position $\mathbf{X}_i$ to a new, candidate position $\mathbf{X}_c$. Then, according to RP's acceptance rule ensuring the detailed balance condition, accept $\mathbf{X}_c$ as the new sample $\mathbf{X}_{i+1}$ with probability $\min\{\pi(\mathbf{X}_c \,|\, \mathbf{Y})/\pi(\mathbf{X}_i \,|\, \mathbf{Y})\}$. If the proposed $\mathbf{X}_c$ is rejected, $\mathbf{X}_{i+1} = \mathbf{X}_i$; if accepted, $\mathbf{X}_{i+1} = \mathbf{X}_c$. $\mathbf{X}_{i+1}$ then serves as the starting point for a subsequent proposal procedure. All accepted paths are then used in the evaluation of $\langle G(\mathbf{X}) \rangle$.

The convergence of this procedure to the desired $\pi(\mathbf{X}|\mathbf{Y})$ may be slow especially when $\mathbf{X}$ is high dimensional. One reason is that it treats $\mathbf{X}$ as a random walker in making proposals, often resulting in low acceptance rates and rather limited movements in the path space. The asymptotic distribution $\pi(\mathbf{X} \,|\, \mathbf{Y})$ may, in practice, be unreachable.

Enforcing detailed balance, i.e., "reversibility," on a Monte Carlo proposal is a sufficient but not a necessary condition for convergence [45–47].

### B. Hamiltonian Monte Carlo

#### 1. General idea of HMC-like sampling

Hamiltonian Monte Carlo (HMC) [16–18] strikes out in a new direction. It adds to the path $\mathbf{X}$ an additional set of variables, which we call $\mathbf{P}$, and identifies a search in $(\mathbf{X}, \mathbf{P})$ space that preserves some invariants of the rules for moving about that space. If, for example, the motion in the extended space preserved $\mathbf{P} \cdot \mathbf{P} + \mathbf{X} \cdot \mathbf{X}$, then moving about the space by performing rotations would allow us to move interesting distances in $(\mathbf{X}, \mathbf{P})$ space by simply rotating the variables in the high-dimensional extended space while staying on the $\mathbf{P} \cdot \mathbf{P} + \mathbf{X} \cdot \mathbf{X} = \text{const}$ surface. Making this motion in the extended space may allow large moves in $\mathbf{X}$ space with large probabilities of acceptance. If this were the case, one might be sampling $\exp[-A(\mathbf{X})]$ much more efficiently than detailed balance-preserved random walking as in the RP procedures.

As introduced in [16] the added conjugate variables are selected from the well-explored examples of the *canonical momentum* familiar to physicists and thoroughly analyzed for two centuries. By choosing the additional variables $\mathbf{P}$ to be canonical conjugates of $\mathbf{X}$, one can use the rules of classical mechanics to move around in $(\mathbf{X}, \mathbf{P})$ space and preserve the underlying symplectic structure.

If the movement is labeled by a scalar $s$ which we call "time," then the rule

$$\frac{d}{ds}\begin{pmatrix}\mathbf{X}(s) \\ \mathbf{P}(s)\end{pmatrix} = \begin{pmatrix}\mathbf{0} & \mathbf{I} \\ -\mathbf{I} & \mathbf{0}\end{pmatrix}\begin{pmatrix}\nabla_{\mathbf{X}} H(\mathbf{X}(s), \mathbf{P}(s)) \\ \nabla_{\mathbf{P}} H(\mathbf{X}(s), \mathbf{P}(s))\end{pmatrix} \quad (12)$$

preserves the value of the scalar function $H(\mathbf{X}, \mathbf{P})$ as well as volumes in $(\mathbf{X}, \mathbf{P})$ space and a collection of other quantities known as the Poincaré invariants [48,49].

For HMC, in addition to the original target distribution $\pi(\mathbf{X} \,|\, \mathbf{Y}) \propto \exp[-A(\mathbf{X})]$, one needs to choose an arbitrary distribution $\pi(\mathbf{P})$ for the canonical momenta $\mathbf{P}$ in order to fully specify the joint HMC target $\pi(\mathbf{X}, \mathbf{P} \,|\, \mathbf{Y})$. The Hamiltonian can be written as

$$H(\mathbf{X}, \mathbf{P}) = -\log[\pi(\mathbf{X}, \mathbf{P} \,|\, \mathbf{Y})] = A(\mathbf{X}) + h(\mathbf{P}),$$

where $h(\mathbf{P})$ is $-\log[\pi(\mathbf{P})]$ up to an additive constant. The expected value of $G(\mathbf{X})$ is now

$$\int d\mathbf{X}\, d\mathbf{P}\, G(\mathbf{X})\, \pi(\mathbf{P})\, \pi(\mathbf{X} \,|\, \mathbf{Y}) = \frac{\int d\mathbf{X}\, d\mathbf{P}\, G(\mathbf{X})\, e^{-H(\mathbf{X},\mathbf{P})}}{\int d\mathbf{X}\, d\mathbf{P}\, e^{-H(\mathbf{X},\mathbf{P})}}$$

$$= \langle G(\mathbf{X}) \rangle. \quad (13)$$

Equation (13) tells us that the expected values, the goals of SDA and machine learning, are unchanged under HMC for an arbitrary choice of the canonical momenta.

The combination of symmetry-preserving movements in $(\mathbf{X}, \mathbf{P})$ and invariance in the expected values is appealing. These properties make higher acceptance rates and faster mixing for sampling $\pi(\mathbf{X} \,|\, \mathbf{Y})$ possible. Other symmetries in $(\mathbf{X}, \mathbf{P})$ requiring different rules for generating motions might work equally well. HMC should be viewed as one of a class of approaches within this overall idea [50,51]. It should also be noted that a high acceptance rate itself does not guarantee efficiency: combining high acceptance with fast exploration of phase space is a goal of all Monte Carlo methods.

One must distinguish the use of the scalar label $s$ in HMC from the time labels $t$ and $\tau$ as described in Sec. II. $s$ is just a label to track movements in the enlarged space. Actual times $t$ and $\tau$ are labels used in identifying the path of model dynamical variables $\mathbf{x}(t)$ and the data $\mathbf{y}(\tau)$ through an observation window.

#### 2. HMC itself

As discussed above, the HMC procedure [16–18] doubles the dimension of path space $\mathbf{X}$, introducing a physics-motivated but arbitrarily chosen canonical momentum $\mathbf{P}$ associated with the path-space position $\mathbf{X}$.

While certainly not required, the choice for the "kinetic energy," $h(\mathbf{P}) = \mathbf{P} \cdot \mathbf{P}/2$, is convenient and the resulting distribution $\exp[-h(\mathbf{P})]$ is Gaussian. As pointed out by Kadakia [52], small modifications of this standard choice to include higher order than quadratic polynomials of $\mathbf{P}$ can introduce chaotic behavior and substantial mixing in motions generated by $H(\mathbf{X}, \mathbf{P})$ and thus may avoid some potential problems in the implementation of a quadratic choice for $h(\mathbf{P})$.

We proceed with the quadratic choice and the Hamiltonian becomes

$$H(\mathbf{X}, \mathbf{P}) = A(\mathbf{X}) + \frac{\mathbf{P} \cdot \mathbf{P}}{2}. \quad (14)$$

Why does one want to sample in the even larger $(\mathbf{X}, \mathbf{P})$ space given that $\mathbf{X}$ is already high dimensional? The answer lies in the invariance properties of Hamiltonian dynamics. If HMC proposals are made using integration of Eq. (12) from "time" 0 to $s$ with the choice of $H$ as in Eq. (14), namely,

$$\frac{d}{ds}\begin{pmatrix}\mathbf{X}(s) \\ \mathbf{P}(s)\end{pmatrix} = \begin{pmatrix}\mathbf{P}(s) \\ -\nabla A(\mathbf{X}(s))\end{pmatrix}, \quad (15)$$

then $H(\mathbf{X}, \mathbf{P})$ is conserved along the canonical phase-space orbit labeled by $s$. As discussed, many other quantities are preserved. Among them, HMC makes use of the conservation of phase-space volume, i.e., $d\mathbf{X}(s)\, d\mathbf{P}(s) = d\mathbf{X}(0)\, d\mathbf{P}(0)$.

The core of HMC is to propose $(\mathbf{X}(s), -\mathbf{P}(s))$ starting from $(\mathbf{X}(0), \mathbf{P}(0))$. This is precise when the integration is performed with $s$ taken as a continuous variable, so a Hamiltonian flow is realized and $H(\mathbf{X}, \mathbf{P})$ is conserved. A complete HMC proposal includes a negative sign before $\mathbf{P}(s)$, meaning that we need to flip the momentum *after* the Hamiltonian integration is finished. This flipping is to make the proposal reversible in $s$ and symmetric in $(\mathbf{X}, \mathbf{P})$, thus ensuring detailed balance.

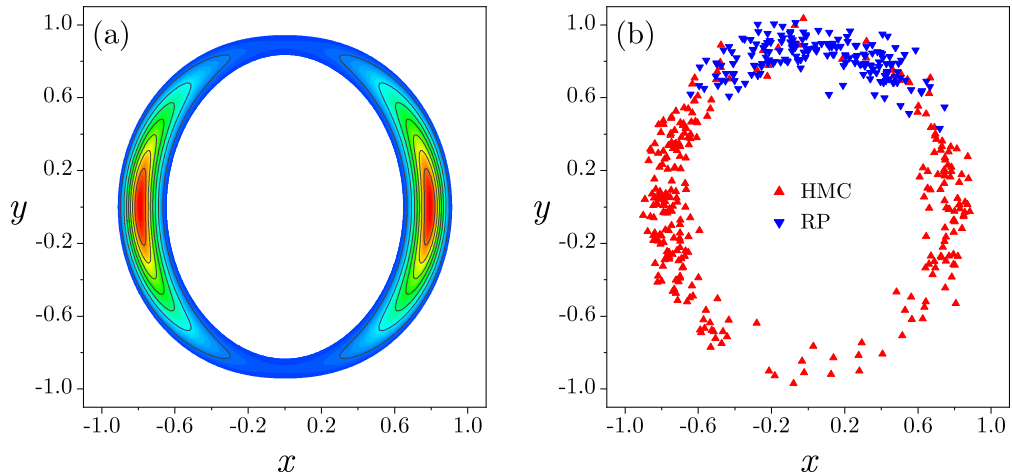

FIG. 3. A comparison of the efficiency of the HMC procedure and the standard Metropolis-Hastings Monte Carlo (RP) procedure. (a) A plot of the target distribution $\pi(x, y) \propto \exp[-(8x^2 + 6y^2 - 5)^2 - 3y^2)]$ within the range $-1 \leqslant x, y \leqslant 1$. (b) The samples generated by RP are shown as blue down-pointing triangles, and the samples generated by HMC are shown as red up-pointing triangles. To arrive at one proposed sample, the discrete Hamiltonian dynamics is simulated using the leapfrog method for $S = 50$ steps with step size $\epsilon = 0.01$. The overall acceptance rate for HMC is 0.998. For the Metropolis algorithm, to arrive at one proposed sample, the $x$ and $y$ directions are perturbed simultaneously by drawing from $\mathcal{N}(0, 0.01)$. The overall acceptance rate for the Metropolis algorithm is 0.56.

In practice, the form of $A(\mathbf{X})$ precludes analytical evaluations, and one must work with discrete $s$ in order to integrate Eq. (15). When we use a symplectic integrator to perform the integration of Hamilton's equations in discrete $s$ [50], a result of Ge and Marsden [53] tells us that we cannot precisely preserve both the symplectic form *and* $H(\mathbf{X}, \mathbf{P})$ the same time. Therefore, we expect that by using a discrete $s$ symplectic integrator, we will find $H(\mathbf{X}(s), \mathbf{P}(s)) \approx H(\mathbf{X}(0), \mathbf{P}(0))$, but not exactly equal. As a consequence, when determining the acceptance or rejection of the proposal $(\mathbf{X}(s), \mathbf{P}(s))$, the acceptance probability can usually be near unity, but not precisely unity, i.e.,

$$\alpha(\mathbf{X}(s), -\mathbf{P}(s) \,|\, \mathbf{X}(0), \mathbf{P}(0))$$
$$= \min\left\{1, \frac{\exp[-H(\mathbf{X}(s), -\mathbf{P}(s))]}{\exp[-H(\mathbf{X}(0), \mathbf{P}(0))]}\right\}$$
$$\approx 1. \qquad (16)$$

The use of $-\mathbf{P}(s)$ can usually be omitted when performing HMC as $\mathbf{P}$ does not enter the calculation of expectations in Eq. (13).

As the parameters $\boldsymbol{\theta}$ in SDA and machine learning are taken to be components of the vector $\mathbf{X}$, they also have conjugate variables $\boldsymbol{\eta}$ included in $\mathbf{P}$. Therefore, HMC provides a principled manner of exploring $\pi(\mathbf{X} \,|\, \mathbf{Y})$ on both state variables and time-independent parameters.

### 3. HMC in practice

When implementing HMC, one must select a symplectic integrator in order to preserve the symplectic invariants [17,50,51], and many choices are available. We have used a fairly common one, the leapfrog symplectic integrator [50,54], which possesses simplicity and accuracy. Using this choice, we select a small step size $\epsilon$ in $s$ to produce a candidate proposal $(\mathbf{X}_c, \mathbf{P}_c)$ by sampling $\exp[-H(\mathbf{X}, \mathbf{P})]$ to select $\mathbf{P}(0)$ and choosing $\mathbf{X}(0)$ to be the last accepted $\mathbf{X}$.

We first move $(\mathbf{X}(0), \mathbf{P}(0))$ to $(\mathbf{X}(\epsilon), \mathbf{P}(\epsilon))$ using the leapfrog symplectic stepping rule,

$$\mathbf{P}(\epsilon/2) = \mathbf{P}(0) - \frac{\epsilon}{2} \nabla A(\mathbf{X}(0)),$$
$$\mathbf{X}(\epsilon) = \mathbf{X}(0) + \epsilon \mathbf{P}(\epsilon/2),$$
$$\mathbf{P}(\epsilon) = \mathbf{P}(\epsilon/2) - \frac{\epsilon}{2} \nabla A(\mathbf{X}(\epsilon)). \qquad (17)$$

Then, starting from $(\mathbf{X}(\epsilon), \mathbf{P}(\epsilon))$, we move to the next point at $(\mathbf{X}(2\epsilon), \mathbf{P}(2\epsilon))$. After proceeding for $S$ steps, we arrive at a candidate proposal $(\mathbf{X}_c, \mathbf{P}_c) = (\mathbf{X}(S\epsilon), -\mathbf{P}(S\epsilon))$. The candidate proposal $(\mathbf{X}_c, \mathbf{P}_c)$ is then accepted or rejected according to Eq. (16).

When using PAHMC (as described below in Sec. V), the HMC procedure is repeated for each value of $R_f$ in Eq. (11). In addition, we choose to perform $N_I$ independent PAHMC calculations in parallel starting from $N_I$ independent initializations.

## IV. HMC vs RP SAMPLING IN A SIMPLE 2D EXAMPLE

We provide a two-dimensional example that illustrates the advantages of sampling the joint distribution $\pi(\mathbf{X}, \mathbf{P})$ using HMC over the standard RP Metropolis-Hastings procedure. In our example, we applied both methods to the distribution $\pi(x, y) \propto \exp[-(8x^2 + 6y^2 - 5)^2 - 3y^2)]$. We started both methods at $(x, y) = (0, 0.8)$. Each of HMC and RP then generates 500 samples, and the last 301 of these samples are retained and displayed in Fig. 3(b). The RP method explores only a small percentage of the distribution while HMC clearly explores most of the distribution in this comparison. The efficiency of HMC compared to RP is clear from this elementary example.

The example in Fig. 3 shows that HMC typically achieves better mixing than the RP method. In addition, it is important to compare the scaling properties of both methods as the dimension of the target distribution increases. With the potential applications to practical problems in mind, it is critical that the sampling process remains efficient in higher dimensions. It has been shown that HMC typically scales better than RP in terms of exploration efficiency. After equilibrium is reached, for a $d$-dimensional target distribution, the computational complexity for a given acceptance rate typically grows as $d^{5/4}$ for HMC [55–57] and $d^2$ for RP [58].

## V. PRECISION ANNEALING METHOD

We now turn to the precision annealing method suggested in [36,38,39]. It is used here to facilitate the search for the global minimum of the action $A(\mathbf{X})$ as we gradually increase the model precision parameter $R_f$. As we will see, this procedure drives the sampling area of HMC to the vicinity of the global minimum in $A(\mathbf{X})$ in Eq. (11) where the path $\mathbf{X}$ is consistent with both the nonlinear model $\mathbf{f}(\mathbf{x}(n), \boldsymbol{\theta})$ and the observations $\mathbf{Y}$. This procedure plays a key role in evaluating the integral in Eq. (9). Because we only have partial observations of the model state, the unobserved degrees of freedom in the state variable $\mathbf{X}$ are completely uninformed upon initialization. If we impose the model to a high precision at the outset, the convoluted model nonlinearities may prevent a sampling scheme from finding the desired minimum in $A(\mathbf{X})$, therefore, the dominant contribution to the integral in Eq. (9) may not be identified.

### A. Initialization in the path space

Our strategy in this paper is to vary the precision parameter $R_f$ that enforces the model error term in $A(\mathbf{X})$. When $R_f = 0$, the model is completely unresolved and the path space is highly degenerate for the unobserved degrees of freedom. In the opposite limit for the hyperparameter $R_f$ where $R_f \to \infty$, the model $\mathbf{f}(\mathbf{x}(n), \boldsymbol{\theta})$ becomes deterministic and provides nonlinear constraints on the minimization of the action. Moving $R_f$ from very small to quite large by a slow increase permits us to start our Monte Carlo procedure near the maximum of $\pi(\mathbf{X} \mid \mathbf{Y})$ and remain there.

We now present the initialization procedure at $R_f = 0$. In this case, the standard model action becomes a quadratic form as

$$A(\mathbf{X}) = \sum_{k=0}^{F} \sum_{\ell=1}^{L} \frac{R_m}{2(F+1)} [x_\ell(k) - y_\ell(k)]^2, \quad (18)$$

which apparently has its (degenerate) global minimum at $x_\ell(k) = y_\ell(k)$ for $k = 0, \ldots, F$ and $\ell = 1, \ldots, L$. Recall that a path $\mathbf{X}$ includes the set of $D$-dimensional state variables $\{\mathbf{x}(0), \ldots, \mathbf{x}(M)\}$ and the time-independent model parameters $\boldsymbol{\theta}$. Equation (18) only specifies the initial values for $L$ out of $D$ dimensions in the state variables and leaves the other components free and undetermined.

To complete our choice of initial path, we now integrate the model forward through $M$ discrete times. $\mathbf{x}(0)$ is initialized such that $x_a(0) = y_a(0)$ for $a = 1, \ldots, L$ and $x_a(0)$ is drawn from a uniform distribution covering the dynamical range of the model for $a = L+1, \ldots, D$. The parameters $\boldsymbol{\theta}$ are also drawn from an appropriate uniform distribution. The model is then integrated forward in time, with the observed dimensions in $\mathbf{x}(t_m)$ replaced by $y_\ell$ at each time where we have a measurement $t_k = \tau_k$ for $k = 0, \ldots, F-1$:

$$\begin{aligned} x_a(k+1) &= y_a(k+1), \quad a = 1, \ldots, L \\ x_a(k+1) &= f_a(\mathbf{x}(k), \boldsymbol{\theta}), \quad a = L+1, \ldots, D. \end{aligned} \quad (19)$$

This completes our construction of an initial path $\mathbf{X}_{\text{init}}$. The freedom in choosing $\mathbf{x}(0)$ and $\boldsymbol{\theta}$ gives us flexibility to generate multiple such initial paths $\mathbf{X}_{\text{init}}^{(q)}$, $q = 1, \ldots, N_I$, at $R_f = 0$. These are retained for future use.

### B. Precision annealing as $R_f$ is increased

Precision annealing arises from the idea that if we choose an initial path for a Monte Carlo or a variational optimization [36–39] at the global minimum for small $R_f$ as we slowly increase $R_f$, we will stay in a region of phase space where our sampling will arrive at the smallest minimum of the action, for a variational calculation, or sample the neighborhood of the maximum of $\pi(\mathbf{X}, \mathbf{P})$, for a Monte Carlo search.

We adopted the following annealing schedule for $R_f$:

$$R_f = R_{f_0} \alpha^\beta \quad (20)$$

with $\alpha > 1$ and $\beta = 0, 1, \ldots, \beta_{\max}$. $R_{f_0}$ should be small. A choice of $\alpha$ near unity leads to the slow increase in $R_f$ as $\beta$ increases, introducing the nonlinearity of the model in an adiabatic manner. At each $R_f$ value, $N_I$ MC calculations are performed starting from the solution generated by the procedure in the previous $R_f$.

At $\beta = 0$, we select, without repeats, each of the $N_I$ initial paths $\mathbf{X}_{\text{init}}^{(q)}$ identified at $R_f = 0$ to be an initial condition. We then start to sample from the HMC joint distribution $\pi(\mathbf{X}, \mathbf{P}) \propto \exp[-h(\mathbf{P}, \mathbf{X}) - A(\mathbf{X})]$ which has $R_{f_0}$ as the hyperparameter. We collect $N_{\beta=0}$ sampled paths, denoted by $\mathbf{X}_{\beta=0, j}^{(q)}$, with $j = 1, \ldots, N_{\beta=0}$, that are generated along the Markov chain. We take the sample mean of each of the $N_I$ Markov chains developed so far,

$$\overline{\mathbf{X}}_{\beta=0} = \frac{1}{N_{\beta=0}} \sum_{j=1}^{N_{\beta=0}} \mathbf{X}_{\beta=0, j}, \quad (21)$$

as the initial path for the next $\beta$ value. Since $N_I$ calculations are done independently and in parallel, this results in a set of $N_I$ averages $\overline{\mathbf{X}}_{\beta=0}^{(q)}$, with $q = 1, \ldots, N_I$.

Next, we move to $\beta = 1$. We have $N_I$ selections of initial conditions, i.e., $\overline{\mathbf{X}}_{\beta=0}^{(q)}$, with $q = 1, \ldots, N_I$, available for our HMC evaluations. For each initial condition we collect $N_{\beta=1}$ sampled paths and form the $N_I$ means for each:

$$\overline{\mathbf{X}}_{\beta=1}^{(q)} = \frac{1}{N_{\beta=1}} \sum_{j=1}^{N_{\beta=1}} \mathbf{X}_{\beta=1, j}^{(q)}, \quad (22)$$

We continue this procedure until we reach a useful $\beta_{\max}$. By plotting two quantities versus $R_f$ or $\beta = \log_\alpha(R_f/R_{f_0})$ we will have insight on how to select this.

First, we should make a plot of the action determined by each of the $N_I$ paths at each $R_f$ as a function of $\beta$. We find that the action becomes independent of $\beta$ as the precision of the model increases. The origin of this is seen in a second graphic where we plot the model error term in the action,

$$\sum_{m=0}^{M-1} \sum_{a=1}^{D} \frac{R_f(a)}{2M} [x_a(n+1) - f_a(\mathbf{x}(n), \boldsymbol{\theta})]^2, \quad (23)$$

as a function of $\beta$.

If the model is consistent with the data, the model error will rapidly decrease to a small value as $\beta$ increases. The action is the sum of the model error, the measurement error, and the initial condition, as the model error tends numerically to a small number, the action levels out to the sum of the measurement error and the initial condition, which are independent of $R_f$. Indeed, this second sum both provides a lower bound to the action and indicates whether the selected model is "right." This independence of the action with respect to $\beta$ will also depend in SDA on the number of measurements $L$ at each time observations are made.

As a result of these MC calculations, the expected value of functions $G(\mathbf{X})$ in the original space, Eq. (9), will be estimated as

$$\langle G(\mathbf{X}) \rangle \approx \frac{1}{N_I} \sum_{q=1}^{N_I} G(\overline{\mathbf{X}}_{\beta_{\max}}^{(q)}) \quad (24)$$

since the precision annealing procedure locates the paths that dominate $\pi(\mathbf{X} | \mathbf{Y}) \propto \exp[-A(\mathbf{X})]$ [36–39]. In the examples we will present below, we have chosen $N_I = 30$ for HMC and $N_I = 50$ for RP Monte Carlo and taken $\beta_{\max}$ in the range 30–50.

## VI. ANALYSIS OF THE Lorenz96 MODEL WITH $D = 20$

The Lorenz96 model [21] is an ideal test bed for validating new methods for nonlinear dynamical systems. Its dynamical equations are simple and elegant, yet its chaotic nature makes it hard to achieve good prediction results.

The model has a $D$-dimensional state variable $\mathbf{x}(t) = (x_1(t), \ldots, x_D(t))$ satisfying

$$\frac{dx_a(t)}{dt} = x_{a-1}(t)[x_{a+1}(t) - x_{a-2}(t)] - x_a(t) + \nu, \quad (25)$$

for $a = 1, \ldots, D$. In Eq. (25), $x_{-1}(t) = x_{D-1}(t)$, $x_0(t) = x_D(t)$, and $x_1(t) = x_{D+1}(t)$. $\nu$ is a constant forcing term; the solutions of these equations for $D \geqslant 4$ are chaotic when $\nu \geqslant 8.0$ or so. We report on calculations with $D = 20$ with $\nu_{\text{true}} = 8.17$. During the calculations and the predictions, $\nu_{\text{true}}$ is not given to the method.

It is useful to note that we do not simply fit a single parameter $\nu$ in the series of numerical examples in this section. Instead, for each case, we need to estimate *all* the $D(M + 1) + 1$ degrees of freedom in the path $\mathbf{X}$. A detailed explanation on this matter is given in Sec. VIII A. As a result, the present problem is still high dimensional albeit having only one parameter $\nu$.

### A. Using PAHMC

We now apply the HMC methods we have outlined to the analysis of the Lorenz96 model with $D = 20$. We are performing a "twin experiment" [1] here in which we solve Eq. (25) for $\mathbf{x}(t)$ beginning with an arbitrary initial value $\mathbf{x}(0)$ and using a time step $\Delta t = 0.025$. To each of the $D$ time series we add Gaussian noise with standard deviation $\sigma = 0.4$ throughout the observation window $\{t_0, \ldots, t_{\text{final}}\}$ which contains $M + 1$ time steps spaced at $\Delta t$. These serve as our data. In the twin experiments, we know the (noisy) time series $\mathbf{y}(t)$ and the value of the forcing parameter $\nu$ so that we are able to validate the results of the proposed method.

We select $L < D$ components of our data set as observations and seek to estimate all $D$ state variables within the observation window as well as the forcing $\nu$. Assuming at each model time step $t = t_k$ we have an observation $\mathbf{y}(k)$, i.e., $n_k = k$ in Eq. (6) for $k = 1, \ldots, M$, the corresponding action is

$$A(\mathbf{X}) = \sum_{k=0}^{F} \sum_{\ell=1}^{L} \frac{R_m}{2(F+1)} [x_\ell(\tau_k) - y_\ell(\tau_k)]^2$$
$$+ \sum_{m=0}^{M-1} \sum_{a=1}^{D} \frac{R_f}{2M} [x_a(m+1) - f_a(\mathbf{x}(m), \nu)]^2. \quad (26)$$

The first term on the right in Eq. (11) is the measurement error, and the second, the model error. The trapezoidal rule is used to discretize Eq. (25), such that

$$f_a(\mathbf{x}(m), \nu) = x_a(m) + \frac{\Delta t}{2} [\mathcal{F}_a(\mathbf{x}(m+1), \nu) + \mathcal{F}_a(\mathbf{x}(m), \nu)], \quad (27)$$

where $\mathcal{F}_a(\mathbf{x}, \nu)$ is the model vector field.

In using Monte Carlo methods, we sample the conditional probability distribution $\pi(\mathbf{X} | \mathbf{Y}) \propto \exp[-A(\mathbf{X})]$ so we are able to estimate both an average value for the path $\mathbf{X}$ as well as variations about this mean. Both will be presented. We will discuss results for $L = 7, 8, 10$, and 12.

#### 1. HMC with $L = 7$

We begin with $L = 7$ observed noisy time series out of 20 generated as our data set. We used the seven data series $y_\ell(m)$ with $\ell \in \{1, 4, 7, 10, 13, 16, 19\}$ as the observations. In Fig. 4(a), we display the action levels as a function of $R_f$ for $L = 7$. In this figure, many action levels are present, corresponding to many peaks in the probability $\pi(\mathbf{X} | \mathbf{Y}) \propto \exp[-A(\mathbf{X})]$ identified. At large $R_f$ values, the $N_I = 30$ action levels start to plateau. However, they are all well above the expected action value which is the measurement error piece of the action.

This is an indication that though HMC has located some paths that agree with the Lorenz96 model in Eq. (25), it fails to transfer information from the observations $\mathbf{Y}$ to the model due to an insufficient number of observations at each $\tau_k$. We expect that as the number of observations $L$ increases, more information will become available to the precision annealing HMC method and that the structure of the action levels will change.

The inadequacy of using only seven observations at each measurement time, $L = 7$, is shown in the failure of the model error to significantly decrease with increasing $R_f$. This is seen in Fig. 4(b). Similarly, in the estimation of the model forcing parameter shown in Fig. 4(c), we see substantial inaccuracies.

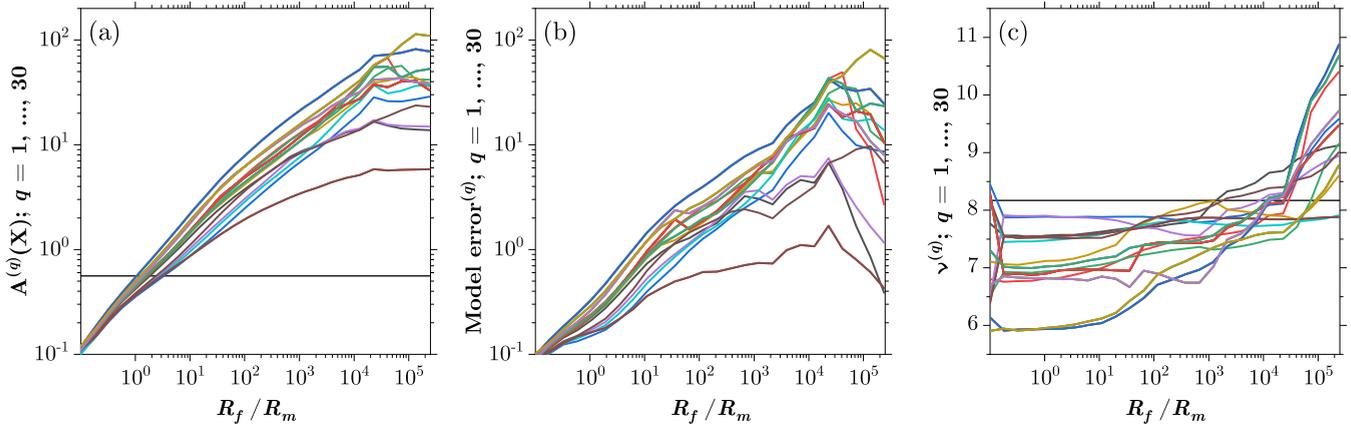

FIG. 4. Calculation results using HMC for the Lorenz96 model with $D = 20$ and $L = 7$. Each step in the precision annealing procedure is associated with one value of $R_f$. We perform $N_I = 30$ independent calculations starting at $N_I$ different $(\mathbf{X}(0), \mathbf{P}(0))$. (a) Action levels versus $R_f$. Because $N_I$ calculations are performed at each $R_f$, many action curves are displayed. All the action levels are above the anticipated value, which is the value of the measurement error term of the total action, shown as the solid black line. (b) The model errors in Eq. (11) as a function of $R_f$. The model errors for the seven observed state variables remain large for all $R_f$. (c) The estimations of the Lorenz96 model forcing parameter $\nu$ as a function of $R_f$. The true forcing parameter is $\nu = 8.17$ (solid black line). At $L = 7$, the estimates of the forcing parameter are not accurate.

Further in the predictions for an observed state variable $x_7(t)$ [Fig. 5(a)] and for an unobserved state variable $x_{14}(t)$ [Fig. 5(b)], we see many errors in both the estimation window $0 \leqslant t \leqslant 5$ and the prediction window $5 < t \leqslant 11$.

### 2. HMC with $L = 8$

In Fig. 6(a), we present the $L = 8$ action levels. Here, 7 out of 30 action levels split from the rest at an early stage in the precision annealing process and coincide with the anticipated action level for the measurement error term. This implies that the transfer of information has been successful. Indeed, as we will see in the figures that follow, the estimations in the model parameter $\nu$ as well as the predictions beyond the observation window show that the paths with lowest action levels yield consistent results. In addition, Fig. 6(b) shows the model error calculations for $L = 8$. It is clear from there that the 7 lowest action levels are numerically dominated by their respective measurement errors instead of model errors.

Figure 6(c) shows $N_I$ estimations of the Lorenz96 forcing parameter $\nu$. The true value is $\nu = 8.17$. If we combine this plot with Fig. 6(a), we can conclude that although the method has located some path with high model precision, some of the paths (23 out of 30) differ from the observations. The "wrong" models have been identified because there are not enough measurements $L$ each time an observation is made, leading to errors in the estimated forcing parameter.

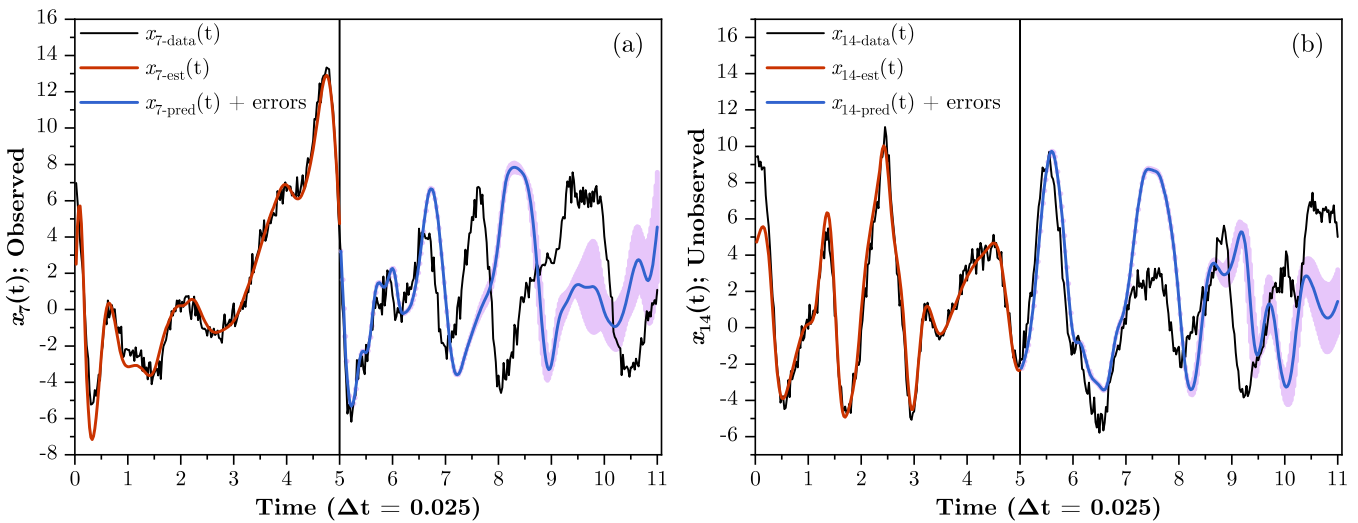

FIG. 5. Prediction results using HMC for the Lorenz96 model with $D = 20$ and $L = 7$. (a) Display of results for an observed dynamical variable $x_7(t)$. The noisy data lie in the time interval $0 \leqslant t \leqslant 11$. The estimated values in the observation window $0 \leqslant t < 5$ are shown in red. The predicted values are shown in blue and the prediction errors are shown in magenta in $5 < t \leqslant 11$. (b) Display of results for an unobserved dynamical variable $x_{14}(t)$. The true data, with noise added, span the full $0 \leqslant t \leqslant 11$ interval. The estimated values in the observation window $0 \leqslant t < 5$ are shown in red. The predicted values, with errors (in magenta), are shown in blue within the interval of $5 < t \leqslant 11$. For the unobserved variables, the data within $0 \leqslant t \leqslant 11$ are unavailable to the method.

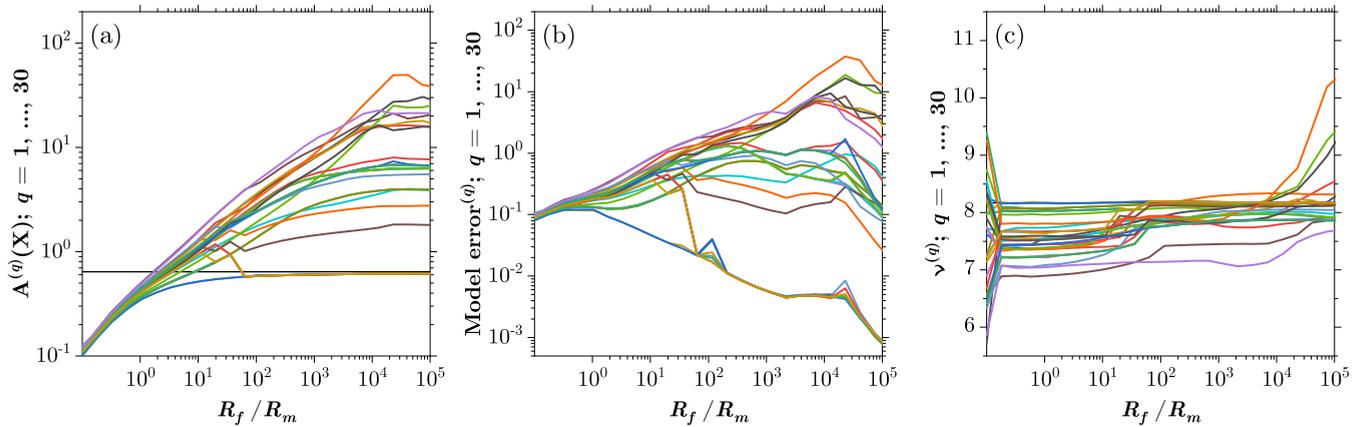

FIG. 6. Calculation results using HMC for the Lorenz96 model with $D = 20$ and $L = 8$. Each step in the precision annealing procedure is associated with one value of $R_f$. We perform $N_I = 30$ independent calculations starting at $N_I$ different $(\mathbf{X}(0), \mathbf{P}(0))$. (a) Action levels versus $R_f$. 23 of 30 action levels are above the anticipated value, shown as the solid black line. (b) The model errors in Eq. (11) as a function of $R_f$. At smaller $R_f$ values, the model error dominates the action, indicating that the HMC calculations have not yet found a path that agrees with the model described by Eq. (27). At the final stage, the model error starts to decrease exponentially and the paths proposed by HMC start to agree with the model. (c) The estimations of the forcing parameter $\nu$ as a function of $R_f$. The true forcing parameter is $\nu = 8.17$.

Until this point, we have examined the indicators, i.e., action level, model error, estimated forcing, representing the quality of the transfer of information. All of these are inferred from the calculations happening in the observation window $0 \leqslant t \leqslant 5$ and from the data $\mathbf{Y}$ available to the PAHMC method. To validate our information transfer method, we need to take advantage of the twin experiment setting. We will compare the estimated unobserved state variables with the data generated in the twin experiment. We will also predict forward in time beyond the observation window and compare the predictions with the actual data.

Figure 7(a) shows the estimate of an observed variable $x_9(t)$ within the estimation window $0 \leqslant t \leqslant 5$ as well as its prediction within the window $5 < t \leqslant 11$, for Lorenz96 $L = 8$. We can see that the estimation agrees quite well with the data. The prediction, with the HMC calculation of errors shown, agrees with the data in the prediction window for some time (about 4 inverse Lyapunov times) and eventually diverges due to the chaotic nature of the Lorenz96 system at $\nu = 8.17$. It is worth noticing that the errors in the prediction are quite small in earlier stages. This is because the PAHMC procedure has accurately estimated all of the observed and unobserved state variables as well as the forcing parameter $\nu$.

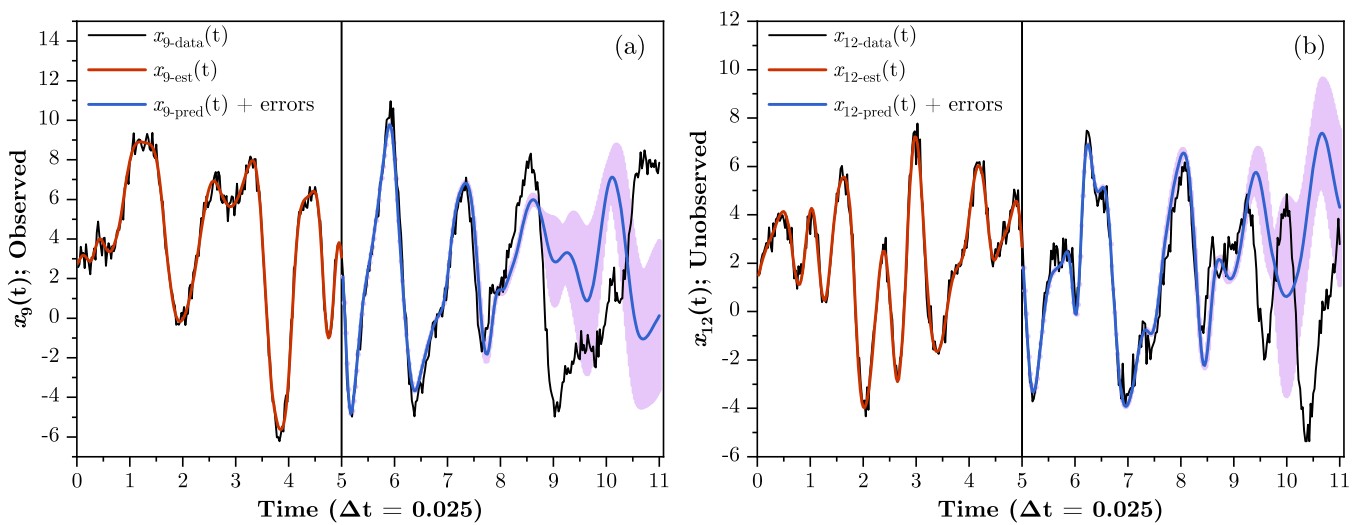

FIG. 7. Prediction results using HMC for the Lorenz96 model with $D = 20$ and $L = 8$. (a) Display of results for an observed dynamical variable $x_9(t)$. The noisy data lie in the time interval $0 \leqslant t \leqslant 11$. The estimated values in the observation window $0 \leqslant t < 5$ are shown in red. The predicted values are shown in blue and the prediction errors are shown in magenta in $5 < t \leqslant 11$. (b) Display of results for an unobserved dynamical variable $x_{12}(t)$. The true data, with noise added, span the full $0 \leqslant t \leqslant 11$ interval. The estimated values in the observation window $0 \leqslant t < 5$ are shown in red. The predicted values, with errors (in magenta), are shown in blue within the interval of $5 < t \leqslant 11$. For the unobserved variables, the data within $0 \leqslant t \leqslant 11$ are unavailable to the method.

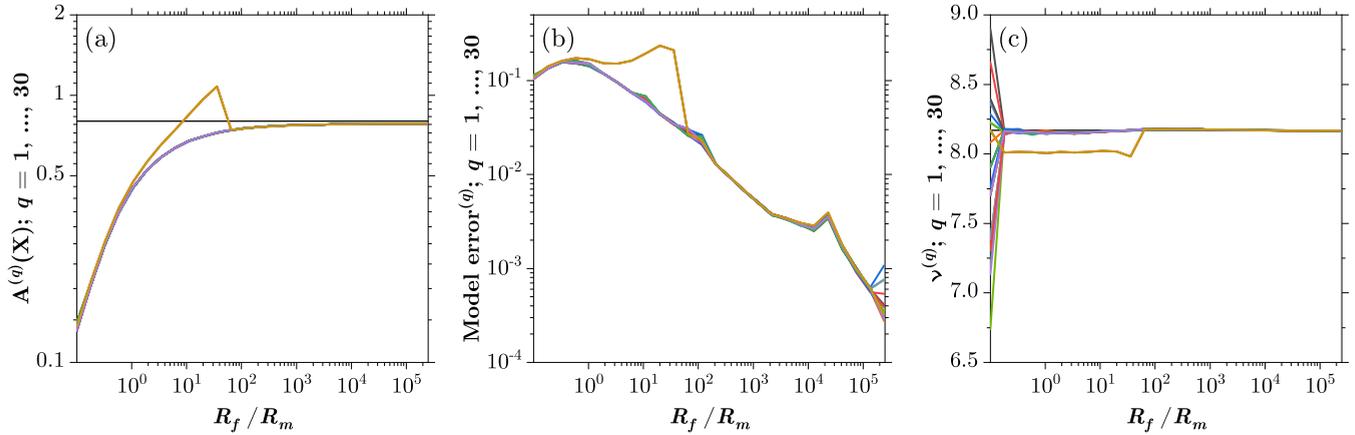

FIG. 8. Calculation results using HMC for the Lorenz96 model with $D = 20$ and $L = 10$. Each step in the precision annealing procedure is associated with one value of $R_f$. We perform $N_I = 30$ independent calculations starting at $N_I$ different $(\mathbf{X}(0), \mathbf{P}(0))$. (a) Action levels at each $R_f$. All these levels agree with the anticipated value, shown as the solid black line. (b) The model errors as a function of $R_f$. At smaller $R_f$ values, the model error dominates the action. After that, the model error rapidly decreases as $R_f$ grows, and the measurement error term dominates the action, becoming essentially independent of $R_f$. (c) The estimations of the forcing parameter $\nu$ at each value of $R_f$. The true forcing parameter in the twin experiments is $\nu = 8.17$. The correct forcing parameter quickly emerges from the random initialization.

Figure 7(b) presents the same information, but for an *unobserved* variable $x_{12}(t)$. In this case, the data for $x_{12}(t)$ are not presented to the PAHMC method in the observation window, yet it still achieves high accuracy in both estimation and prediction. This is a strong indication that the transfer of information from $\mathbf{Y}$ to Lorenz96 has been successful for $L = 8$ observations. The nonlinearity of the model is at work here transferring information from observed to unobserved state variables.

#### 3. HMC with $L = 10$

If we further increase the number of observations to $L = 10$, we see that all the action levels plateau almost at the anticipated expected value, as shown in Fig. 8(a). This indicates that the information is sufficient for the method to locate the global minimum of the action. Also, in Fig. 8(b), the rapidly decaying model errors show that the dynamics $\mathbf{x}(m + 1) = \mathbf{f}(\mathbf{x}(m), \nu)$ is enforced strictly in the large-$R_f$ regime. Figure 8(c) shows the estimated forcing parameter for $L = 10$ as a function of $R_f$. The 30 estimated forcing parameters quickly converge to the true value of 8.17, which indicates that the correct model has been found by all the $N_I = 30$ HMC calculations. Conclusions similar to the $L = 8$ case can be drawn from the prediction results in Figs. 9(a) for the observed variable $x_{17}(t)$ and 9(b) for the unobserved state variable $x_{14}(t)$.

#### 4. HMC with $L = 12$

Increasing $L$ to 12 produces no additional information about properties of the source of the data or structure of the action. In Fig. 10 we see the results for the action, the model error, and the Lorenz96 forcing parameter for the $N_I = 30$ initial paths used in the calculations. We forgo showing that observed and unobserved state values are estimated quite well both in the observation window $0 \leqslant t \leqslant 5]$ and the prediction window $5 < t \leqslant 11$, only noting that the earlier calculations have already informed us that $L$ near 8 or 9 is enough to capture the properties of the noisy data.

### B. Using random proposal Monte Carlo

Our numerical calculations are "twin experiments" in which for a selected $D$ we choose $\mathbf{x}(t_0) = \mathbf{x}(0)$ and using a time step $\Delta t = 0.025$ generate solutions $\mathbf{x}(t)$ over an observation window $[t_0, t_{\text{final}}]$ in which $t_0 \leqslant t \leqslant t_0 + M \Delta t = t_{\text{final}}$. To each $x_a(t)$, we add Gaussian noise with mean zero and variance $\sigma^2 = 0.25$; these now comprise our library of observed data. We then select $L \leqslant D$ of these noisy data, and form the action

$$A(\mathbf{X}) = \sum_{k=0}^{F} \sum_{\ell=1}^{L} \frac{R_m}{2} [x_\ell(\tau_k) - y_\ell(\tau_k)]^2$$
$$+ \sum_{m=0}^{M-1} \sum_{a=1}^{D} \frac{R_f}{2} [x_a(m+1) - f_a(\mathbf{x}(m), \nu)]^2. \quad (28)$$

Our calculations were performed with the following choices: $D = 20$, $\alpha = 1.4$, $R_{f0} = 4.0$, $R_m = 4.0$, $N_I = 50$, $\Delta t = 0.025$, and two choices of $L$, 5 and 12.

#### 1. RP with $L = 5$

In Fig. 11, we display the action levels as a function of $R_f$ at $L = 5$. We can see that RP identifies many action levels, corresponding to many peaks in the conditional probability distribution $P(\mathbf{X} | \mathbf{Y}) \propto \exp[-A(\mathbf{X})]$ in Eq. (28). From $R_f = 2 \times 10^4$, we see one level moving away from the collection of larger action levels as $\beta$ increases. However, no action level has become essentially independent of $R_f$, suggesting that the accuracy with which the model is enforced remains too small. We expect that as the number of measurements at each $\tau_k$ is increased, more information about the phase-space instabilities will be passed from the data to the model and that the structure of the action level plot will change.

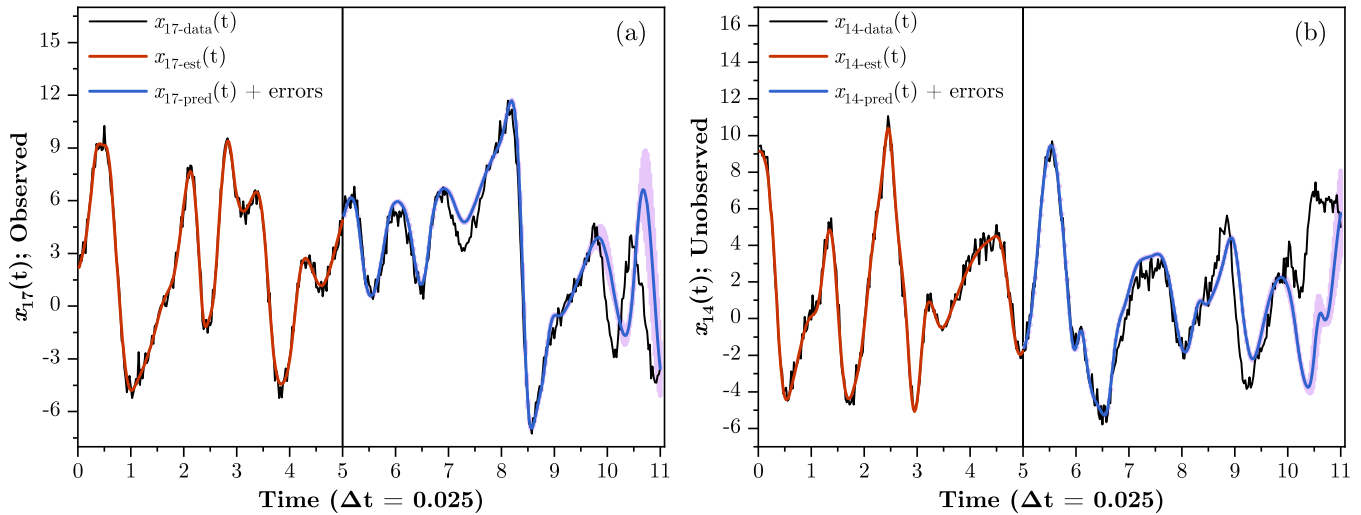

FIG. 9. Prediction results using HMC for the Lorenz96 model with $D = 20$ and $L = 10$. (a) Display of results for an observed dynamical variable $x_{17}(t)$. The data, with noise added, span the full $0 \leqslant t \leqslant 11$ interval. The estimated values in the observation window $0 \leqslant t < 5$ are shown in red. The predicted values, with errors (in magenta), are shown in blue within the interval of $5 < t \leqslant 11$. (b) Display of results for an unobserved dynamical variable $x_{14}(t)$. The true data, with noise added, span the full $0 \leqslant t \leqslant 11$ interval. The estimated values in the observation window $0 \leqslant t < 5$ are shown in red. The predicted values, with errors (in magenta), are shown in blue within the interval of $5 < t \leqslant 11$. For the unobserved variables, the data within $0 \leqslant t \leqslant 11$ are unavailable to the method.

#### 2. RP with L = 12

In Fig. 12, we now display the action levels and its components, the measurement errors and the model errors, when $L = 12$. Here the behavior of the action levels is quite different. The model error decreases over a large range of $R_f$ until the numerical stability of the evaluation of this term is reduced as small errors in $\mathbf{x}(m+1) - \mathbf{f}(\mathbf{x}(m), \nu)$ are magnified by large values of $R_f$. As this result appears, the action for each of the $N_I$ paths at each $\beta$ levels off, becoming essentially independent of $R_f$, and matches the measurement error, as it must do for consistency [36–39].

Until this point, we have examined outcomes of the RP precision annealing estimation procedure. All of the state variables, *observed* and *unobserved*, as well as the forcing parameter were reported over the observation window $0 \leqslant t \leqslant 5$. In a "twin experiment" as here, we have generated the data by solving a known dynamical equation and adding noise to the output of the $D = 20$ times series with a known value of $\nu$. The point of a twin experiment is to test the method of transfer of information in SDA. As we have $D - L$ unobserved state variables at each $L$, and an unobserved parameter $\nu$, the only tool to determine how well the estimation procedure has done in its task is to predict for $t > 5$ into a prediction window where no information from observations is passed back from the model. We now examine how well the estimation has been performed by predicting both an observed and an unobserved time series among the $D$ available.

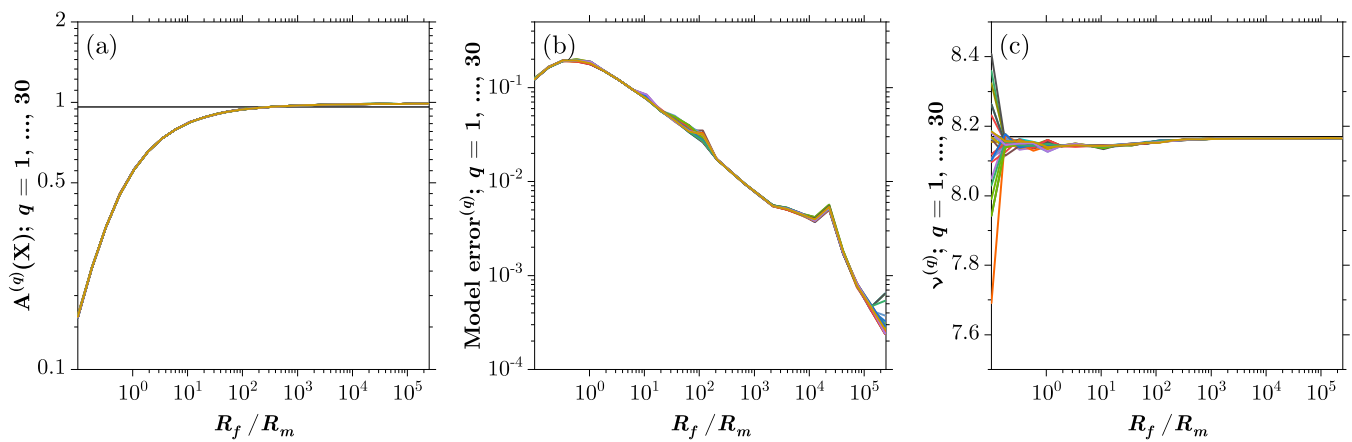

FIG. 10. Calculation results using HMC for the Lorenz96 model with $D = 20$ and $L = 12$. Each step in the precision annealing procedure is associated with one value of $R_f$. We perform $N_I = 30$ independent calculations starting at $N_I$ different $(\mathbf{X}(0), \mathbf{P}(0))$. (a) Action values at each $R_f$. All the action levels quickly plateau at the anticipated level as $R_f$ grows. (b) The model errors as a function of $R_f$. At smaller $R_f$ values, the model error dominates the action. After that, the model error rapidly decreases as $R_f$ grows, and the measurement error term dominates the action, becoming essentially independent of $R_f$. This indicates that the precision annealing procedure has successfully located the path that agrees well with the measurements and the model. (c) The estimations of the forcing parameter $\nu$ at each value of $R_f$. The true forcing parameter is $\nu = 8.17$.

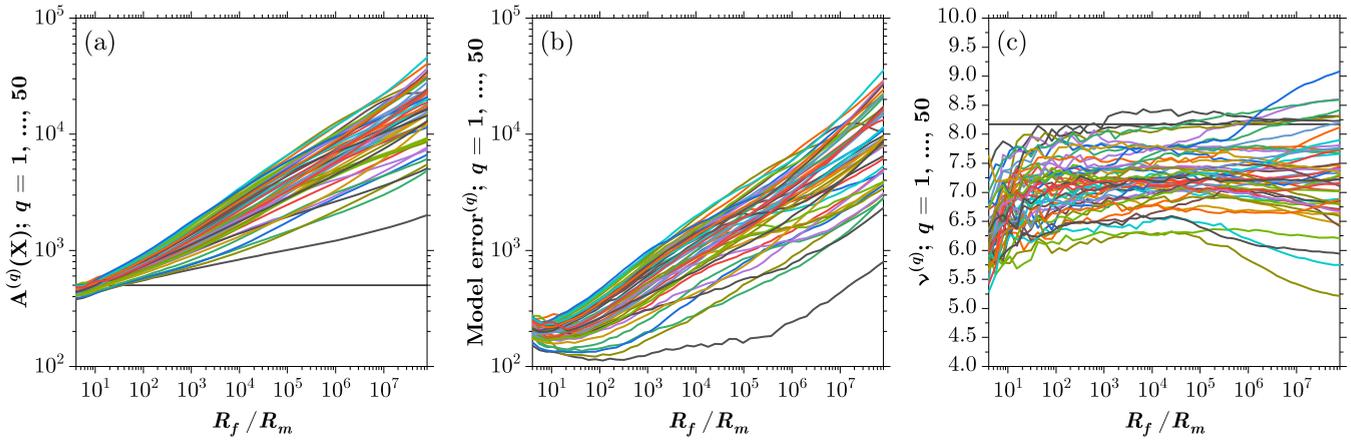

FIG. 11. Calculation results using random proposal Monte Carlo for the Lorenz96 model with $D = 20$ and $L = 5$. We perform precision annealing with RP starting with $N_I = 50$ initial paths. (a) The values of the actions as in Eq. (28). The actions are evaluated at $\beta = \log_\alpha(R_f/R_m)$ where $\alpha = 1.4$ and $R_m = 4.0$. Displayed here are the 50 action values at each $R_f$. These actions are evaluated along the expected path resulting from the sampled paths generated during the Metropolis-Hastings procedures from each of the $N_I$ initial paths. (b) The values of the model errors. (c) The values of the forcing parameter.

We already see from Fig. 12(c) that the input value of $\nu = 8.17$ has not been estimated very accurately. The apparent bias in this parameter estimation has also been seen in an earlier Monte Carlo twin experiment [35,59]. However, notice that HMC can accurately estimate the parameter $\nu$ without an apparent bias as shown in Fig. 10.

Figure 13(a) shows the observed model variable $x_2(t)$ for the Lorenz96 model with total dimensions $D = 20$, observed dimensions $L = 12$, and $\Delta t = 0.025$. The noisy data from solutions of the model equations are from the observed variables $\{1, 2, 4, 6, 7, 9, 11, 12, 14, 16, 17, 19\}$. The estimation of $x_2(t)$ during the observation window using RP to transfer information from the data to the model is shown in red, and the prediction using all the estimated states of the model, $\mathbf{x}(t = 5)$, and the estimated model parameter, is shown in green $\mathbf{x}(t \geqslant 5)$. Our knowledge of this dynamical system [59] indicates that the largest global Lyapunov exponent is approximately 1.2 in the time units indicated by $\Delta t$. The deviation of the predicted trajectory $x_2(t)$ from the data near $t \approx 6.0$ is consistent with the accuracy of the estimated state $\mathbf{x}(t)$ and this Lyapunov exponent.

Figure 13(b) shows the unobserved model variable $x_{20}(t)$ for the same Lorenz96 model. The noisy data from solutions of the model equations are from the observed variables $\{1, 2, 4, 6, 7, 9, 11, 12, 14, 16, 17, 19\}$. The estimation of $x_{20}(t)$ during the observation window using RP to transfer information from the data to the model is shown in red, and the prediction using all the estimated states of the model, $\mathbf{x}(t = 5)$, and the estimated model parameter, is shown in blue $\mathbf{x}(t \geqslant 5)$. Our knowledge of this dynamical system [59] indicates that the largest global Lyapunov exponent is approximately 1.2 in the time units indicated by $\Delta t$. The deviation of the predicted trajectory $x_{20}(t)$ from $t \approx 6.4$ is consistent with the accuracy of the estimated state $\mathbf{x}(t)$ and this Lyapunov exponent.

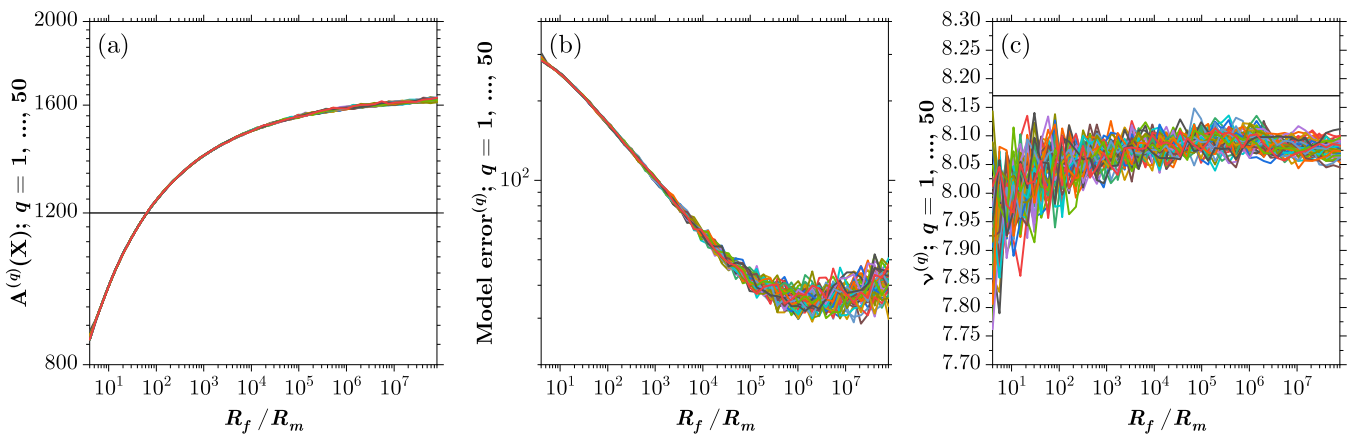

FIG. 12. Calculation results using random proposal Monte Carlo for the Lorenz96 model with $D = 20$ and $L = 12$. We perform precision annealing with RP starting with $N_I = 50$ initial paths. (a) The values of the actions as in Eq. (28). The actions are evaluated as a function of $\beta = \log_\alpha(R_f/R_m)$ where $\alpha = 1.4$ and $R_m = 4.0$. These actions are evaluated along the expected path resulting from the sampled paths generated during the Metropolis-Hastings procedures from each of the $N_I$ initial paths. (b) The values of the model errors. (c) The values of the forcing parameter.

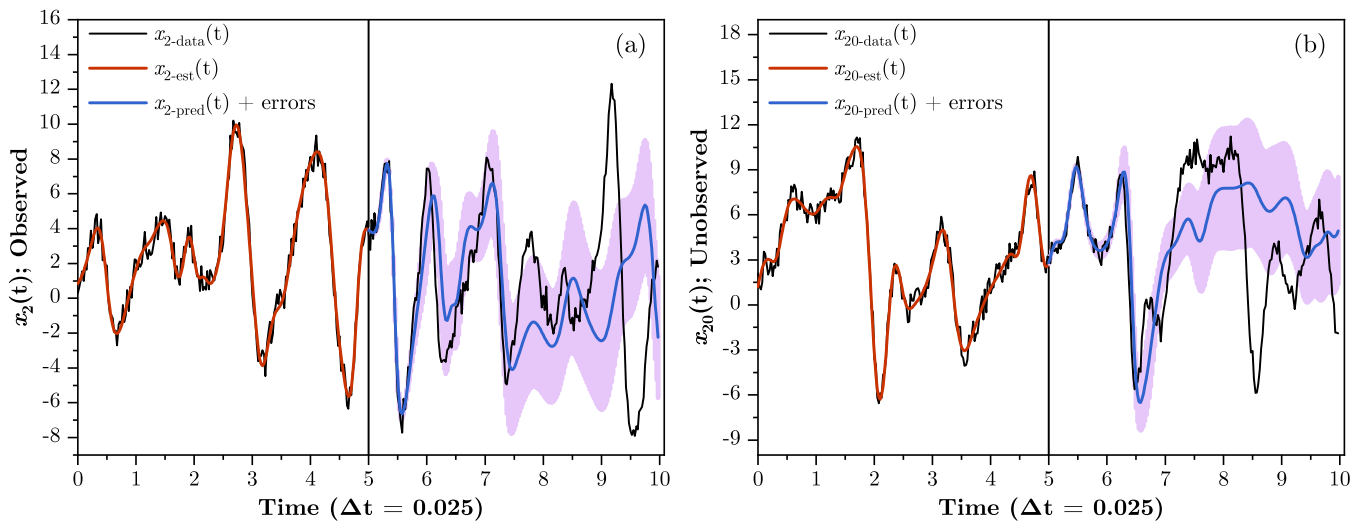

FIG. 13. Prediction results using random proposal Monte Carlo for the Lorenz96 model with $D = 20$ and $L = 12$. The predictions use the values of $\mathbf{x}(t = 5)$ for the full estimated state at the end of the observation window as well as the parameter $\nu$ estimated in the precision annealing procedure. (a) The observed dynamical variable $x_2(t)$ for the time interval $0 \leqslant t \leqslant 10$. In black is the full set of data. In red is the estimated $x_2(t)$ over the observation window $0 \leqslant t \leqslant 5$, and in blue is the predicted $x_2(t)$ over the prediction window $5 < t \leqslant 10$. The RMS prediction errors are in magenta. (b) The unobserved dynamical variable $x_{20}(t)$. In black is the full set of data. In red is the estimated $x_{20}(t)$ over the observation window, and in blue is the predicted $x_{20}(t)$ over the prediction window. The RMS prediction errors are in magenta.

## VII. COMPUTATIONAL COMPLEXITY OF PAHMC

It is important to consider the computational complexity of the proposed method for utilization in a practical task, and it is vital that the method achieves good scaling properties when applied to large systems. Here, we give an estimate on the scaling as the system size increases. We first consider the complexity of an elementary step of HMC, which is that of a one-iteration leapfrog simulation. This requires evaluating $\nabla A(\mathbf{X})$ once and therefore has dominant time complexity of $\mathcal{O}(DM)$, where, as usual, $D$ is the dimension of the underlying dynamical system and $M$ is the number of discrete time steps in the observation window. As the dimension of the path $\mathbf{X}$ grows, the number of leapfrog steps in one HMC proposal needs to be adjusted accordingly to reach a nearly independent point. In fact, it has been noted [17,55,56] that the computation typically grows as the 5/4 power of the dimensions of the model, given a constant acceptance rate.

As a result, the time complexity of the proposed method in this paper is $\mathcal{O}[(DM)^{5/4}]$. Apparently, there exists a multiplicative constant, independent of $D$ and $M$, that is determined by the choices of the number of samples $N_\beta$ and the number of precision annealing steps $\beta_{\max}$.

The mixing time itself is hard to estimate, yet it has been empirically tested in many cases that HMC achieves faster mixing than other well-known Monte Carlo methods. An elegant modification of the term $h(\mathbf{P}, \mathbf{X})$ in the Hamiltonian has been made by Kadakia [52] that may improve the mixing rate.

It is useful to provide a benchmark as to the actual computation time for a real system. To this end, we report the time for the Lorenz96 system considered in this paper. In Lorenz96, we are dealing with a 4000-dimensional system for which $D = 20$ and $M = 200$ (ignoring the additional dimension introduced by the model parameter $\nu$). To obtain one HMC proposal $\mathbf{X}_{\text{prop}}$, we simulate the discrete Hamiltonian dynamics given in Eq. (17) for $S = 50$ steps with step size $\epsilon = 10^{-3}$. In an implementation in MATLAB, the computation time is around 0.02 s. We then make $N_\beta = 10^3$ proposals for every $\beta$ up to $\beta_{\max} = 30$. As a result, the entire calculation takes about 10 minutes to finish.

To provide a direct comparison to the RP HMC method, we have run the same twin experiments for the Lorenz96 system, $D = 20$ and $M = 200$. For every $\beta$ up to $\beta_{\max} = 50$, we make 6000 perturbations on $\mathbf{X}$ within each $\beta$ value: this means we make $2.4 \times 10^7$ proposals for each $\beta$, given that we only perturb one component of $\mathbf{X}$. Implemented in C, the calculation took about 4 h to complete. The computation time is much longer than that of the HMC method, since the RP method requires more proposals due to inefficient sampling of $\exp[-A(\mathbf{X})]$.

Aside from the above analyses of the $\mathcal{O}[(DM)^{5/4}]$ rule and the actual computation time, we also point out that the precision annealing HMC method could gain a significant speedup if run in parallel, provided that the action is constructed as Eq. (11). We will again use Lorenz96 to illustrate this principle, but it should be noted that the only change in Eq. (11) for a different system would be the choice for the vector field $f_a(\mathbf{x}(m), \boldsymbol{\theta})$. We will focus on the evaluation of $\nabla A(\mathbf{X})$ since this is the only calculation that scales directly with $D$ and $M$.

First, for the $D$ vector field $\mathcal{F}_a(\mathbf{x}(m), \nu)$ in Eq. (25) and a discretization rule (in time) given by Eq. (27), the Jacobian for $t = t_m$ in the observation window is given as

$$\begin{aligned}
\mathcal{J}_{ij}(m) &= \frac{\partial \mathcal{F}_i(\mathbf{x}(m), \nu)}{\partial x_j(m)} \\
&= \delta_{i-1,j}[x_{i+1}(m) - x_{i-2}(m)] \\
&\quad + (\delta_{i+1,j} - \delta_{i-2,j})x_{i-1}(m) - \delta_{ij}
\end{aligned} \quad (29)$$

with $i, j = 1, \ldots, D$. The time-consuming part in calculating the $D$ components in $\nabla A(\mathbf{X})$ corresponding to $t = t_m$ is to multiply the Jacobian $\mathcal{J}(m)$ with a $D$ vector. This can be fully parallelized to have a $D$ times speedup.

Second, to complete the calculation of $\nabla A(\mathbf{X})$, we need to repeat the above process for each time mark $m$ from 1 to $M$. These repetitions are independent and can therefore be fully parallelized to have an $M$ times speedup.

The good potential of parallelization discussed above makes precision annealing HMC especially suitable for very large dynamical systems in terms of both the dimensionality of the vector field and the length of the observation window.

## VIII. REMARKS, SUMMARY, AND DISCUSSION

### A. Remarks on state-space representation

We have defined the *path* $\mathbf{X}$ in Sec. II, and it is a high-dimensional quantity by its nature. Here, we discuss the necessity of using this state-space representation for our task of transferring information. State-space representations are widely adopted in many fields such as numerical weather prediction and control theory. In this work, Eq. (4) establishes the dynamical model in continuous time. Since we always work in discrete time for any model that is not exactly solvable, we arrive at Eq. (5), which has $x_a(m+1) = f_a(\mathbf{x}(m), \boldsymbol{\theta})$, where $\mathbf{x}(m) = (x_1(m), \ldots, x_D(m))$ is the state vector at time $m$. Recall the definition of $\mathbf{X}$, there are in total $\mathcal{T} \equiv D(M+1) + N_p$ degrees of freedom in $\mathbf{X}$.

Now, given that we can only observe $L$ out of $D$ dimensions of the dynamics, and, to simplify discussion, assuming that we can observe the system at each time $m = 0, \ldots, M$, our (noisy) data can then be denoted as $\mathbf{Y} = \{\mathbf{y}(0), \ldots, \mathbf{y}(M)\}$, where each $\mathbf{y}$ is an $L$-dimensional vector. There are $L(M+1)$ degrees of freedom in $\mathbf{Y}$. There are clearly $\mathcal{K} \equiv (D-L)(M+1) + N_p$ degrees of freedom missing, and these $\mathcal{K}$ degrees of freedom make up the "unobserved" portion of $\mathbf{X}$ and need to be estimated.

In fact, not only the $\mathcal{K}$ but also all $\mathcal{T}$ degrees of freedom in $\mathbf{X}$ should be estimated in all cases. Consequently, even if we had only one parameter $\theta$ in the model (such as in the case of the Lorenz96 model), we may still have thousands of variables left for estimation, which is far more challenging than fitting only a single parameter.

As noted in Sec. II A, the generic data set $\mathbf{Y}$ contains only partial information about $\mathbf{X}$. In addition, even though Eq. (5) is deterministic, the data resulted from observing the dynamical process are noisy. This usually means that $y_a(m) = x_a(m) + \eta$ for $a = 1, \ldots, L$, where $\eta \sim \mathcal{N}(0, \sigma)$. These two facts suggest that every component of $\mathbf{X}$ needs to be estimated.

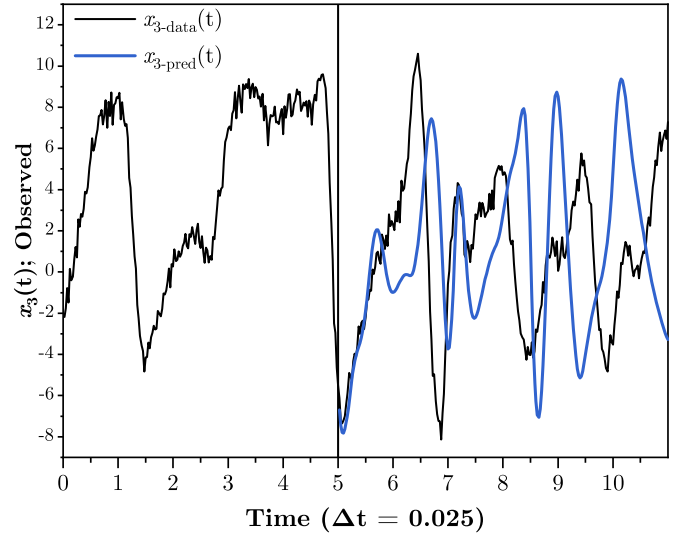

FIG. 14. Predictions for an observed variable $x_3(t)$ for the Lorenz96 model with $D = 20$, $L = 20$, and $\theta_{\text{true}} \equiv \nu_{\text{true}} = 8.17$ known. No effort was made to estimate the path $\mathbf{X}$, and the predictions were produced by integrating forward from $\mathbf{y}(M)$. As a result, the predicted $x_3(t)$ does not match the data in the prediction window.

Even if we could somehow observe all $D$ out of $D$ dimensions (i.e., $\mathcal{K} = N_p$), we still cannot avoid estimating the $D(M+1)$ state variables in $\mathbf{X}$. This is because the noise in $\mathbf{Y}$ would produce a huge model error in the action, thus preventing any method from locating the optimal $\boldsymbol{\theta}$. Suppose in an even more unrealistic scenario we are given the true value of $\boldsymbol{\theta}$ and a full observation ($L = D$) of $\mathbf{X}$, we still cannot avoid estimating the full $\mathbf{X}$, so we cannot use only $\{\mathbf{Y}, \boldsymbol{\theta}_{\text{true}}\}$ to make predictions beyond the observation window.

Figure 14 explains the above fact through the Lorenz96 system where the $D = 20$ dimensions are fully observed and $\nu_{\text{true}} = 8.17$ is given. Nevertheless, starting from the beginning of the prediction window at $t = 5.0$, the prediction for $x_3(t)$ quickly deviates from the truth. Comparing this with the results in Sec. VI (which always have $L < D$ and $\nu_{\text{true}}$ unknown), we can conclude that all the components of $\mathbf{X}$ should be estimated before making predictions.

Another clarifying remark is on the difference between the space spanned by $\{\mathbf{x}(0), \boldsymbol{\theta}\}$ and the higher-dimensional space spanned by $\mathbf{X}$. Since we work with a deterministic model for which $\mathbf{x}(m)$ *could* be obtained by forward integration from $\mathbf{x}(0)$ and $\boldsymbol{\theta}$, can we avoid the $\mathcal{T}$-dimensional space by only working in the $(D + N_p)$-dimensional space spanned by $\{\mathbf{x}(0), \boldsymbol{\theta}\}$? The answer is *No*.

If we do work in the $\{\mathbf{x}(0), \boldsymbol{\theta}\}$ space, we are forced to impose the equality in the dynamical model $x_a(m+1) = f_a(\mathbf{x}(m), \boldsymbol{\theta})$ from the beginning. Geometrically, this means moving on a $(D + N_p)$-dimensional nonconvex surface hoping to locate the abrupt decline in the action that corresponds to the optimal set of $\mathbf{x}(0)$ and $\boldsymbol{\theta}$. There is no means to achieve this.

To conclude, we have shown that one needs to work with *all* the $\mathcal{T} = D(M+1) + N_p$ variables in **X** in order to achieve the transfer of information from the data **Y**. For example, for the Lorenz96 application discussed in Sec. VI, $N_p = 1$ and $\mathcal{T} = 20 \times 200 + 1 = 4001$. As a result, the general problem considered in this paper is by its nature high dimensional and is therefore more challenging than parameter fitting as in the usual statistical physics context.

### B. Summary and discussion

We have presented a systematic method to transfer information from observations **Y** of the degrees of freedom **X** of a nonlinear dynamical system to a model $\mathbf{x}(m) \to \mathbf{x}(m+1) = \mathbf{f}(\mathbf{x}(m), \boldsymbol{\theta})$ of the processes in the system. The observations are noisy, and the model has errors, so we must address the conditional probability distribution $\pi(\mathbf{X}|\mathbf{Y}) \propto \exp[-A(\mathbf{X})]$, conditioned on the observations, as the quantity of our interest. Using this probability, the expected value of any function of **X**, $G(\mathbf{X})$, can be expressed as

$$\langle G(\mathbf{X}) \rangle = \frac{\int d\mathbf{X}\, G(\mathbf{X}) e^{-A(\mathbf{X})}}{\int d\mathbf{X}\, e^{-A(\mathbf{X})}}. \tag{30}$$

The path through the observation window $\mathbf{X} \equiv \{\mathbf{x}(0), \dots \mathbf{x}(M), \boldsymbol{\theta}\}$ is the collection of the model states at the discrete times in that window along with the time-independent model parameters. The approximate numerical evaluation of expected value integrals is the central focus of this paper. We have shown [7,15] this to be the case both in SDA and in neural networks for supervised machine learning.

The hallmark of success in this information transfer process is that one can accurately predict the future states of the dynamical model for some time window, a prediction or generalization interval, beyond the observation window, given the final estimated state $\mathbf{x}(M)$ and parameters $\boldsymbol{\theta}$ at the termination of the observation window. Before being able to predict, one must first complete the model $\mathbf{f}(\mathbf{x}(m), \boldsymbol{\theta})$ by estimating all the components in **X**. The observed components in **X** must also be estimated because of the noise in the data **Y**.

In the use of the SDA methods described in this paper when analyzing physical or biophysical systems, the estimation of the *observed* state variables in the dynamical model $\mathbf{f}(\mathbf{x}(m), \boldsymbol{\theta})$ may turn out to be as important as the estimation of the *unobserved* state variables. In numerical weather prediction, for example, good estimates of the full model state of the earth system provide information not necessarily available in the observations themselves. It is the connection of observed and unobserved state variables within the dynamical model that permits this feature.

This work develops a systematic way of locating the minima of the action $A(\mathbf{X})$ in path, **X**, space, as well as efficiently sampling from the distribution $\pi(\mathbf{X}|\mathbf{Y}) \propto \exp[-A(\mathbf{X})]$ in order to evaluate the path integral. This accurate sampling of $\pi(\mathbf{X}|\mathbf{Y})$ permits estimation of the errors in prediction in addition to expected paths.

We reported on the use of two well-developed Monte Carlo sampling methods in this paper. One is the "hybrid" MC (or Hamiltonian MC) method [16–18] and the other is the familiar Metropolis-Hastings (MH) MC approach where random proposals for moving about in path space for sampling $\pi(\mathbf{X}|\mathbf{Y})$ are made and evaluated. Both on an instructive two-dimensional example and on richer examples drawn from solutions to the Lorenz 1996 [21] model we found that the HMC approach was preferable.

HMC was introduced as an innovative version of Monte Carlo sampling for high-dimensional probability distributions. The core idea is to achieve high efficiency by introducing additional degrees of freedom into the target distribution and avoid the random-walk behavior of MH procedures that have lower acceptance rates of proposed moves in path space as well a markedly slower exploration of the target distribution $\pi(\mathbf{X}|\mathbf{Y})$.

HMC proceeds by making proposals according to Hamilton's equations for the model state variables **X** and their canonical conjugates **P**. The HMC samplings occur in the enlarged in canonical phase space $(\mathbf{X}, \mathbf{P})$, and the target distribution becomes $\pi(\mathbf{X}, \mathbf{P}) \propto \exp[-H(\mathbf{X}, \mathbf{P})]$ with $H(\mathbf{X}, \mathbf{P}) = h(\mathbf{P}) + A(\mathbf{X})$. A collection of conserved quantities are associated with this method of making proposals. Among them are the phase-space volume and $H(|X, \mathbf{P})$ itself. As a consequence, the overall acceptance rate is close, but not equal, to unity.

For a nonconvex, high-dimensional action $A(\mathbf{X})$, HMC itself does not guarantee we will capture the set of maxima of the target distribution without enlightened guidance to find the minima in the action. In fact, for a random initialization, it is unlikely for an HMC sampler to travel for a long distance in the **X** space. Therefore, accurate approximation of the path integral in Eq. (30) also hinges on locating the global minimum in the action of the form

$$A(\mathbf{X}) = \sum_{k=0}^{F} \sum_{\ell=1}^{L} \frac{R_m}{2(F+1)} [x_\ell(\tau_k) - y_\ell(\tau_k)]^2 + \sum_{m=0}^{M-1} \sum_{a=1}^{D} \frac{R_f}{2M} [x_a(m+1) - f_a(\mathbf{x}(m), \boldsymbol{\theta})]^2.$$

To address this issue, we have extended the precision annealing (PA) procedure developed in the context of variational SDA calculations [36–39] to MC procedures: in this paper to MH and HMC strategies alike.

In PA the precision $R_f$ of the dynamical model is varied gradually from $R_f \approx 0$ to very large values facilitating remaining near the global minimum. At $R_f = 0$, the dynamical model is completely unresolved and the global minimum is easily identified. As one increases $R_f$, MC sampling begins at a position in the path space **X** that is well informed by the samples within previous $R_f$ values. This enables the locating of the global minimum in $A(\mathbf{X})$ to be more and more precise as $R_f$ becomes larger. Eventually, if $R_f \to \infty$, the deterministic version of the model is imposed on the calculation.

We tested our method on the Lorenz96 model, with dimension $D = 20$. We report on results where the number of observations ranges from $L = 7$ to 12. We have shown that, at small $L$, the minima located by PAHMC do not necessarily agree with the observations, though they can be in accordance with the dynamical model. As the number of observations increases, the method takes advantage of the additional information available and successfully locates the desired minimum of the action. The prediction results also showed that the transfer of information to the model has resulted in enhanced predictability of the model system. This can be traced to improvements in the estimation of the full state of the model at the termination of the observation window and to the stabilization of intrinsic instabilities in the chaotic nonlinear model [59].

A direct comparison between HMC and RP implementations of precision annealing Monte Carlo on the Lorenz96 model $D = 20$ shows that PAHMC performs approximately 25 times faster. This does not account for the fact that the RP calculations were done using a compiled code written in C while the HMC code was done using a MATLAB interpreted code. The core algorithms utilized by MC procedures can be performed in parallel using GPUs [36]. Contemporary GPU capability suggests a plausible factor of 1000 or more in the speedup of MC calculations. Using PAHMC on very high-dimensional physical (SDA) or supervised ML tasks seems quite promising.

We started by recalling the equivalence of SDA and machine learning [7]. We finish by noting that the role of $L$ is played in machine learning by the number of input/output pairs presented to the proposed network model during training.

## ACKNOWLEDGMENTS

We acknowledge the suggestions from the anonymous referees. We also thank A. Miller for helpful discussions and feedback on the manuscript.